\documentclass[11pt,a4paper]{article}
\pdfoutput=1

\usepackage{jcappub}
\usepackage{hyperref}
\usepackage{graphicx}
\usepackage{epsfig}
\usepackage{subfigure}
\usepackage{comment}
\usepackage{slashed}
\usepackage{soul}
\usepackage{tabularx}
\usepackage{physics}
\usepackage[normalem]{ulem}
\usepackage{afterpage}
\usepackage{placeins}

\usepackage{array}
\newcolumntype{C}[1]{>{\centering\let\newline\\\arraybackslash\hspace{0pt}}m{#1}}

\newcommand{\be}{\begin{equation}} 
\newcommand{\ee}{\end{equation}}
\newcommand{\bea}{\begin{equation}\begin{aligned}} 
\newcommand{\eea}{\end{aligned}\end{equation}}
\newcommand{\ber}{\begin{eqnarray}}
\newcommand{\ear}{\end{eqnarray}}

\def\lsim{\mathrel{\raise.3ex\hbox{$<$\kern-.75em\lower1ex\hbox{$\sim$}}}}
\def\gsim{\mathrel{\raise.3ex\hbox{$>$\kern-.75em\lower1ex\hbox{$\sim$}}}}

\renewcommand{\tr}{{\rm tr}\,}

\newcommand{\GeV}{{\rm GeV}}

\newcommand{\ie}{{\it i.e.}}
\newcommand{\eg}{{\it e.g.\,}}

\newcommand{\td}{{\rm d}}

\newcommand{\mpl}{M_{\rm P}}

\newcommand{\eps}{\epsilon}

\newcommand{\phib}{\bar{\phi}}
\newcommand{\rhob}{\bar{\rho}}

\newcommand{\tomega}{{\tilde{\omega}}}
\newcommand{\tLambda}{{\tilde{\Lambda}}}

\newcommand{\tq}{{\tilde{q}}}

\newcommand{\mth}{m_{\rm th}}
\newcommand{\fgw}{f_\mathrm{GW}}
\newcommand{\mufull}{\lambda}

\newcommand{\pk}{\text{peak}}
\newcommand{\amp}{\text{amp}}

\begin{document}

\title{The Linear Regime of Tachyonic Preheating}

\author{Niko Koivunen,}
\author{Eemeli Tomberg,}
\author{and Hardi Veerm\"{a}e}
\affiliation{NICPB, R\"avala 10, 10143 Tallinn, Estonia}  

\emailAdd{niko.koivunen@kbfi.ee}
\emailAdd{eemeli.tomberg@kbfi.ee}
\emailAdd{hardi.veermae@cern.ch}

\abstract{Tachyonic preheating is realized when the inflaton repeatedly returns to a convex region of the potential during the post-inflationary oscillating phase. This will induce a strong tachyonic instability and lead to a rapid fragmentation of the coherent field that can complete within a fraction of an $e$-fold. In this paper, we study the linear regime of this process in a model-independent way. To this purpose, we construct simplified models that provide an analytic Floquet theoretic description of mode growth. This approach captures the essential features of well-motivated tachyonic preheating scenarios, including scenarios in which the inflaton is part of a larger scalar multiplet. We show that tachyonic preheating is efficient if the field excursions are sub-Planckian, can produce gravitational waves in the frequency range of current and future gravitational wave interferometers, and can be consistent with any experimentally allowed tensor-to-scalar ratio.}

\maketitle

\section{Introduction} \label{sec:intro}

The inflationary paradigm resolves several outstanding problems in Big Bang cosmology~\cite{Starobinsky:1980te, Starobinsky:1983zz, Guth:1980zm, Linde:1981mu, Albrecht:1982wi, Linde:1983gd, Lyth:1998xn, Planck:2018jri} and successfully explains the primordial density perturbations inferred from the cosmic microwave background (CMB) measurements by the Planck satellite~\cite{Planck:2018jri}. 

During inflation, cosmic expansion is driven by the inflaton's potential energy. This energy must eventually be transferred to Standard Model (SM) particles. This process is called reheating~\cite{Kofman:1994rk, Kofman:1997yn}. The initial non-perturbative phase of reheating is dubbed preheating\footnote{The nomenclature differs in literature. Sometimes the distinction is made between the non-perturbative stage of particle production (preheating) and perturbative inflaton decay that produces the thermal bath of SM particles (reheating).} and can proceed through parametric resonance~\cite{Dolgov:1989us,Traschen:1990sw, Kofman:1994rk, Shtanov:1994ce, Kofman:1997yn, Greene:1997fu} or through a tachyonic instability~\cite{Felder:2000hj, Felder:2001kt,Tomberg:2021bll}. In the first case, the post-inflationary oscillating inflaton induces time-dependence in the masses of the fields it is coupled to. This periodically changing mass leads to explosive particle production through a resonant instability. On the other hand, tachyonic preheating occurs when the squared mass of the fields becomes negative, triggering particle production due to a tachyonic instability.

Tachyonic preheating has been studied in the context of hybrid~\cite{Felder:2000hj, Felder:2001kt,Garcia-Bellido:2001dqy, Copeland:2002ku, Barnaby:2006cq, Garcia-Bellido:2007fiu, Dufaux:2008dn}, hilltop~\cite{Desroche:2005yt, Dufaux:2006ee, Antusch:2015nla, Antusch:2015vna, Ema:2017rkk, Antusch:2017vga},
small field \cite{Brax:2010ai},
and 
plateau inflation~\cite{Rubio:2019ypq, Karam:2020rpa,Karam:2021sno,Tomberg:2021bll}. In single field inflation with a potential that is convex at inflationary scales, tachyonic preheating generally occurs when the Hubble scale at the end of inflation is much smaller than the inflaton's effective mass. To have such a sub-mass scale inflation, the potential should become flatter as the field moves away from the minimum. Such setups are naturally realized in plateau inflation. When the condition for the Hubble scale is satisfied, the field repeatedly returns to the tachyonic plateau region as it oscillates. These scenarios also tend to be in good agreement with the increasingly tightening experimental bounds on inflationary parameters~\cite{Martin:2013nzq}. Flattened plateau potentials are quite common and appear, for instance, 
in alpha-attractor models~\cite{Ferrara:2013rsa, Kallosh:2013yoa, Carrasco:2015rva, Galante:2014ifa, Kallosh:2013hoa,Krajewski:2018moi}, 
and in Palatini formulations with non-minimal gravitational couplings \cite{Bauer:2008zj, Rubio:2018ogq, Enckell:2018hmo, Antoniadis:2018ywb, Tenkanen:2020dge}. Also models in~\cite{Starobinsky:1980te, Salopek:1988qh, Kaiser:1994vs, Bezrukov:2007ep} give rise to flat potentials, but these are incapable of producing tachyonic preheating.

The current study provides a simple analytic model-independent description of the initial linear phase of inflaton fragmentation due to a dominant tachyonic instability. It complements our earlier numerical study~\cite{Tomberg:2021bll}. Since the timescales associated with fragmentation in tachyonic preheating are much shorter than the Hubble time, we can neglect the expansion of space. The coherent background then oscillates with an almost constant frequency and amplitude, and the instability of each mode can be understood using Floquet theory. The modes then follow the flat space equation
\be\label{eq:dphi_eom_flat0}
    \delta\ddot{\phi}_k(t) + \qty(m_{\rm eff}^2(t) + k^2) \delta\phi_k(t) = 0 \, ,
\ee
with a time-dependent effective mass $m_{\rm eff}^2$ that becomes negative during a part of each period. Since $m_{\rm eff}^2$ changes sign, the models considered here contain elements of both parametric resonance and tachyonic instability. Moreover, these situations do not obey the Mathieu or the Lame equation, commonly encountered in discussions of parametric resonance~\cite{Traschen:1990sw, Kofman:1994rk, Shtanov:1994ce, Kofman:1997yn, Greene:1997fu}. We note that similar scenarios involving periodic sign-flips of the effective mass, dubbed the flapping resonance, are encountered in the context of axion models~\cite{Kitajima:2018zco,Fukunaga:2019unq}.

Even in the linear phase, describing fragmentation can be a computationally demanding task, since in the absence of a general analytical solution to \eqref{eq:dphi_eom_flat0} the evolution of each mode must be tracked numerically. For this reason, we propose simple analytically solvable models constructed by approximating $m_{\rm eff}^2$ with simpler time-dependence, where $m_{\rm eff}^2$ is constant and negative on the plateau and positive near the minimum of the potential. We choose the positive part as a delta function or a top-hat like box. These simplified models can be understood as temporal analogues to the Kronig-Penney model~\cite{Kronig:1931} that describes electron conduction in a periodic one-dimensional spatial lattice. The obtained analytic picture illuminates how quantitative features of a given inflation model translate into the spectrum of leading instability bands. In particular, this approach provides model-independent analytic estimates for the characteristics of the instability, such as the fastest growing mode and its growth rate. 

Our approach does not describe processes during the final non-linear stage of preheating, such as the creation of secondary peaks due to rescattering, formation of oscillons or the eventual thermalization of the inflaton particles. These processes have so far been studied using lattice methods~\cite{Lozanov:2016hid, Lozanov:2017hjm, Krajewski:2018moi, Lozanov:2019ylm, Bhoonah:2020oov}. We also do not consider the decay of the fragmented inflaton into the SM thermal bath. Nevertheless, the ideas discussed here may be extended to include some of the aforementioned effects. As an example, we consider preheating when the inflaton is embedded into an $O(N)$ multiplet.

This paper is structured as follows. In section~\ref{sec:general}, we give the general set-up required for tachyonic preheating with an oscillating inflaton and outline the Floquet theoretical basis for perturbation growth. Section~\ref{sec:models} introduces the simplified models for which the perturbation growth is analytically solvable. The structure of instability bands is discussed in detail in section~\ref{sec:band_structure} with the aid of the simplified models as well as a numerically worked out realistic scenario. A multi-field scenario is studied in section~\ref{sec:multi_field} and phenomenological implications of our results are discussed in section~\ref{sec:discussion}. We conclude in section~\ref{sec:concl}. Some technical details are given in the appendix. Throughout this paper we use natural units $\hbar = c = 1$ and set the reduced Planck mass to unity, $\mpl = 1$.

\section{General considerations}
\label{sec:general}

We will study the dynamics of the inflaton $\phi$ arising from the Einstein frame action
\be\label{eq:S}
    S = \int\td^4 x \sqrt{-g} \left[ \frac{1}{2} R - \frac{1}{2} (\partial \phi)^2 - U(\phi) \right] \, ,
\ee
where $g$ denotes the determinant of the metric tensor, $R$ is the Ricci scalar, and $U$ is the scalar potential. In a Friedmann--Lema\^itre--Robertson--Walker spacetime, the equation of motion of the coherent background field $\phib$ and the Friedmann equation are
\be\label{eq:FLRW}
    \ddot{\phib} + 3 H \dot{\phib} + U'(\phib) = 0 \, , \qquad 
    3H^2 = \bar{\rho} \, ,
\ee
respectively. Above, $\bar{\rho} \equiv \dot{\phib}^2/2 + U(\phib)$ is the energy density of the scalar, $H\equiv \dot{a}/a$ is the Hubble parameter, $a$ is the scale factor, and a dot denotes a derivative with respect to the cosmic time. These equations describe both inflation and the post-inflationary oscillations of $\phib$ around the minimum of $U$. However, to transition into the hot Big Bang era, the energy density of the background field then has to be transferred into radiation and, ultimately, into the SM degrees of freedom.

The early stages of this process are characterized by the growth of the scalar perturbations $\delta\phi \equiv \phi - \phib$ due to the time-dependence of the background. This non-perturbative particle production after inflation is called preheating, and it ultimately leads into the fragmentation of the background field. We focus on the early linear stages of this process during which each mode of the perturbed field evolves according to
\be\label{eq:dphi_eom}
    \delta\ddot{\phi}_k + 3H\delta\dot{\phi}_k + \omega_k^2\delta\phi_k = 0 \, ,
    \qquad
    \omega_k^2 \equiv k^2/a^2 + U''(\phib) \, .
\ee
As these perturbations grow, interactions between the modes become increasingly important, and the remaining evolution has to be studied using non-linear methods, such as lattice simulations or thermal field theory~\cite{Lozanov:2016hid, Lozanov:2017hjm, Krajewski:2018moi, Lozanov:2019ylm, Bhoonah:2020oov}. 

\begin{figure}
\begin{center}
	\includegraphics[width=.98\linewidth]{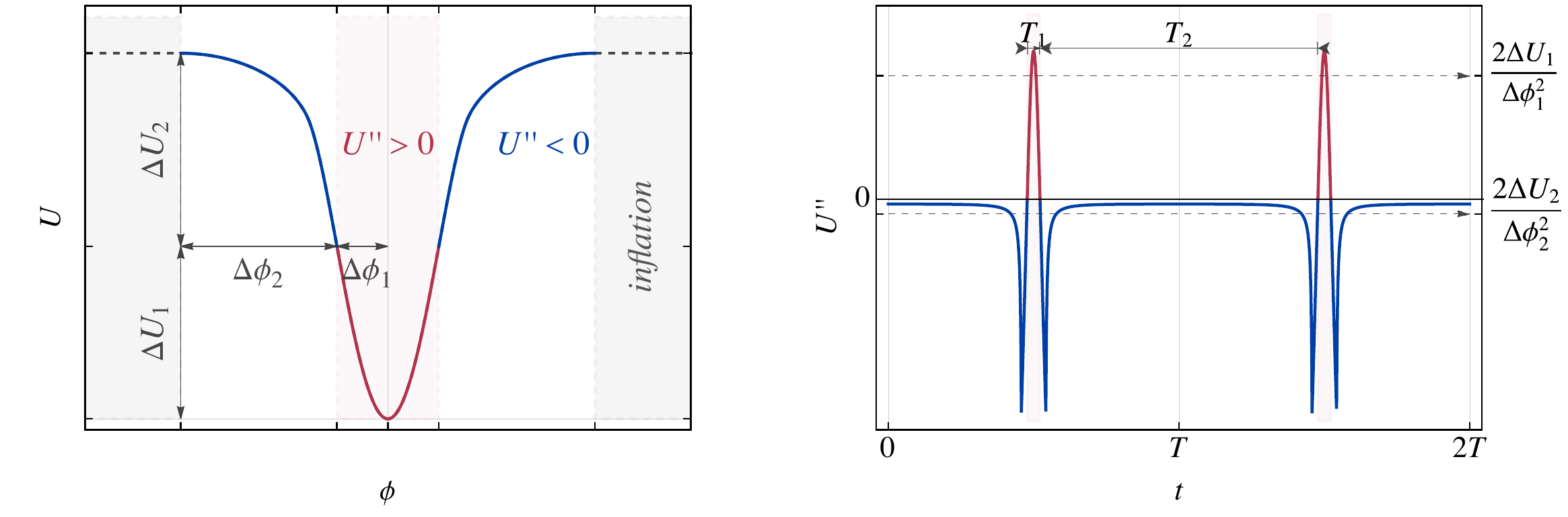}
\end{center}
\vspace{-4mm}
\caption{A schematic of the potential \emph{(left panel)} and the effective squared mass \emph{(right panel)} exhibiting a post-inflationary tachyonic instability. In the blue region, the tachyonic instability is active, while field excursions through the red region induce rapid changes in the mass and contribute to the parametric resonance. The gray region in the left panel correspond to the inflationary regime and there the potential does not have to be specified for the purposes of the current study.}
\label{fig:U_schematic}
\end{figure}

Preheating is more often studied in a context where the inflaton $\phi$ oscillates in a quadratic potential, $U(\phi)\sim\phi^2$, and decays to another scalar field $\chi$ through a coupling of the form $g\phi^2\chi^2$, or the inflaton potential is quartic, $U(\phi)\sim\phi^4$, and the coherent inflaton fragments. The perturbations will then obey a Mathieu or Lame equation, respectively, and the instability responsible for particle production is the parametric resonance~\cite{Traschen:1990sw, Kofman:1994rk, Shtanov:1994ce, Kofman:1997yn, Greene:1997fu}. In both cases, perturbations predominantly grow when $\phi$ crosses zero in the effective broad resonance regime. Instead, we are interested in a different situation, in which $U'' < 0$ is possible away from the minimum of the potential and leads to a tachyonic instability~\cite{Felder:2000hj,Felder:2001kt}. Notice that, around the minimum, we must still have $U''>0$ by construction, so the tachyonic instability cannot be active at all times if the background is to oscillate. 

A schematic of a typical potential that gives rise to tachyonic preheating is shown in the left panel of Fig.~\ref{fig:U_schematic}. The right panel of Fig.~\ref{fig:U_schematic} depicts the corresponding temporal evolution of the effective mass $U''(\phib(t))$. One can observe long stretches in the tachyonic $U'' < 0$ region followed by brief transitions through the non-tachyonic $U'' > 0$ minimum. The height $\Delta U_i$ and width $\Delta \phi_i$ of the corresponding regions give a rough estimate for the effective mass: $U'' \approx 2 \Delta U_1/\Delta \phi_1^2 \equiv m_1^2$ and $U'' \approx -2 \Delta U_2/\Delta \phi_2^2 \equiv -m_2^2$ in the non-tachyonic and the tachyonic region, respectively. The time $T_1$ spent transitioning through the non-tachyonic region is of the order $m_1$ and must satisfy $(1+\Delta U_2/\Delta U_1)^{-1/2} \leq m_1 T_1 /2 \leq (\Delta U_1/\Delta U_2)^{1/2}$, computed from the maximal and minimal velocities in the limit of negligible Hubble friction. Throughout the rest of the paper we assume that the potential is symmetric, so that $U''$ is periodic with period $T = T_1+T_2$, while the background field $\phib$ has a period $2T$.

We do not consider the details of the inflationary epoch, and thus, the shape of the potential beyond the oscillating region is not specified in Fig.~\ref{fig:U_schematic}. Nevertheless, the Hubble rate during the oscillating period is roughly $\Delta U_1 + \Delta U_2 \approx 3 H^2$. In these scenarios, slow-roll ends due to the violation of the second slow-roll condition $U''/U \lesssim 1$ implying that the potential should satisfy $\Delta \phi_2^{-2} \Delta U_2/(\Delta U_1 + \Delta U_2) \gg 1 $. This condition will generally guarantee that Hubble friction is weak in the tachyonic region. For fast preheating, we want this to be true also near the minimum, so we require that the time spent in the non-tachyonic region does not exceed the Hubble time, $T_1 \ll H^{-1}$. Using estimates from the discussion above, we find that this is satisfied when $\Delta \phi_1 (1 + \Delta U_1/\Delta U_2)^{1/2} \ll 1$. This condition is less relevant because, in most cases, the time spent in the tachyonic region is significantly longer, \ie, $T_1 \ll T_2$, since the field velocity is much lower there. In all, we can formulate the sufficient condition 
\be\label{eq:tach_cond}
    (\Delta \phi_1 + \Delta \phi_2) (1 + \Delta U_1/\Delta U_2)^{1/2} \ll 1
\ee
for effective tachyonic preheating. In particular, for $\Delta U_1 \lesssim \Delta U_2$, it is sufficient that the field value at the end of inflation, given roughly by $\Delta \phi_1 + \Delta \phi_2$, is sub-Planckian.

Preheating can be extremely rapid if the field repeatedly returns to the tachyonic region, causing the complete fragmentation of the coherent background within a fraction of an $e$-fold. As the processes under study take place in sub-Hubble timescales, we can neglect cosmic expansion altogether, and the perturbation equation~\eqref{eq:dphi_eom} simplifies to the Hill equation~\cite{1968ZaMM...48R.138R}
\be\label{eq:dphi_eom_flat}
    \delta\ddot{\phi}_k + \omega_k^2\delta\phi_k = 0 \, ,
\ee
while the background \eqref{eq:FLRW} obeys $\ddot{\phib} + U'(\phib) = 0$. We remark that $\omega_k^2(t)$ depends on the energy density of the background $\rhob$, which, even in Minkowski space, will decrease due to the feedback from the perturbation growth. Nevertheless, if $\rhob$ is not damped considerably during a single oscillation, its evolution can be studied in the adiabatic approximation by defining an effective equation of state by taking suitable averages over a single period~\cite{Turner:1983he,Karam:2021sno,Tomberg:2021bll}. Below, we will work in the limit where both the Hubble friction and backreaction can be neglected during an oscillation.

\subsection{Floquet theory}
\label{sec:floquet}

For a periodic $\omega_k^2$, the solutions of Eq.~\eqref{eq:dphi_eom_flat} are governed by Floquet theory, which invites us to look for quasiperiodic solutions of the form
\be\label{eq:exp_growth}
    \delta\phi_k(t+T) = e^{\mufull_k T}\, \delta\phi_k(t) \, ,
\ee
where $\mufull_k$ are the Floquet exponents for the mode $k$. We define the growth rate for the mode $k$ as
\be\label{eq:mu_k}
    \mu_k \equiv \max \Re \mufull_{k}.
\ee
When $\mu_k > 0$, the leading component of the general solution will grow exponentially, which manifests physically as the production of $\phi$-particles. Solving~\eqref{eq:dphi_eom_flat} amounts to finding $\lambda_k$ and, in particular, $\mu_k$, for all values of $k$.

Given two independent solutions $u_1$ and $u_2$ of~\eqref{eq:dphi_eom_flat} we can compute the growth exponents using the monodromy matrix
\be\label{eq:G}
    G = w^{-1}(0)w(T) \, , 
    \qquad \mbox{where} \qquad
    w(t) \equiv
\begin{pmatrix}
    u_1(t) & u_2(t) \\
    \dot{u}_1(t) &  \dot{u}_2(t) \\
\end{pmatrix}
\ee
is the Wronskian matrix. The $G$ is a constant matrix with eigenvalues $e^{\mufull_k T}$, with the two exponents $\mufull_k$ being additive inverses due to the conservation of the Wronskian. Thus, they can be solved as
\be\label{eq:mu_from_trG}
    \mufull_k \, T = \pm {\rm acosh}\left( \frac{1}{2} \tr G \right).
\ee
To find $\mufull_k$ in the numerical examples studied in section~\ref{sec:band_structure}, we solve $u_1$ and $u_2$ from $0$ to $T$ with initial conditions $u_1(0)=\dot{u}_2(0)=1$, $u_2(0)=\dot{u}_1(0)=0$, so $\tr G = u_1(T) + \dot{u}_2(T)$.

An important property of growth rates for inflaton perturbations is that the $k=0$ mode must be stable, that is, $\mu_{k=0}=0$~\cite{Tomberg:2021bll}. In single field inflation, the wavenumber at which the first instability band begins depends on how the period reacts to changes in the energy density of the background field: when $\partial_{\rhob} T > 0$, then the first instability band begins at $k=0$, while for $\partial_{\rhob} T < 0$, there is a stability band beginning at $k=0$~\cite{Tomberg:2021bll}. In detail, for small $k$, the Floquet exponent can be expanded as
\be\label{eq:k_small_growth}
    \mufull_k = -\frac{i\pi}{T} + (k/a) \frac{\sqrt{W \partial_{\rhob} T}} {T} + \mathcal{O}(k/a)^3 \, ,
\ee
and is thus completely determined by the abbreviated action (see also appendix~\ref{app:SON})
\bea \label{eq:W}
    W(\rho) 
&    = 2 \int^{\phi_\amp}_{0} \td \phi \sqrt{2(\rhob - U(\phi))} \, ,
\eea
where $\rhob = U(\phi_\amp)$ is the background energy density and $\phi_\amp$ is the field oscillation amplitude. In terms of the abbreviated action, the duration of a half-oscillation of $\phib$ is $T = \partial_{\rhob} W$. The presently considered models possess an instability band starting at $k=0$, that is, $\mu_k>0$ immediately as $k>0$ entering a wide band comprising of all the modes for which $\omega_k^2$ is negative on the plateau. This situation resembles the more commonly studied broad parametric resonance regime~\cite{Kofman:1994rk}.

\section{Simplified models}
\label{sec:models}

The common characteristics of the potentials supporting tachyonic preheating scenarios suggest that the essential features of this process can be studied by generic simplified models. In the following sections, we construct such models for the growth of linear perturbations in Eq.~\eqref{eq:dphi_eom_flat} by studying different approximations of $\omega_k^2$. As depicted in Fig.~\ref{fig:U_schematic}, in the models of interest the background is characterized by long stretches in the tachyonic regime followed by rapid crossings of the origin where the sign of $\omega_k^2$ is briefly flipped. This behaviour can be qualitatively captured by the ansatz
\be\label{eq:omega_peak_approx}
	\omega_k^2 
	= -\Gamma_k^2 + \sum_{j \in \mathbb{Z}} f(t - jT) \, ,
\ee
where $f(t)$ is a positive $k$-independent function peaked at $t=0$ describing the temporal evolution of $U''$ around origin crossings and\footnote{To include expansion, one must change $k \to k/a$ to account for redshifting of the modes.}
\be \label{eq:gamma0_in_k}
	\Gamma_k^2 = \Gamma_0^2 - k^2 \, ,
\ee
where $\Gamma_0$ is a complex constant.\footnote{Where it matters, we choose $\Re \Gamma_k > 0$. However, the system is invariant under $\Gamma_k\leftrightarrow-\Gamma_k$.} The background mode $k=0$ should be stable by Eq.~\eqref{eq:k_small_growth}. This provides an additional constraint between $\Gamma_0$, $T$, and $f(t)$. If the ansatz \eqref{eq:omega_peak_approx} describes evolution arising from a specific potential, this determines $\Gamma_0$ and $T$, which then depend on $\rhob$ (or $\phi_\amp$).

Although we focus on models exhibiting tachyonic instabilities, by extending our results to an imaginary $\Gamma_0$, our analytic approximations may be adapted to preheating via parametric resonance only. This approach may be useful when the linear regime is not governed by the Mathieu or Lame equation.

In the following, we will consider two separate ans\"atze for $f(t)$: the delta model and its generalization, the box model.

\subsection{Delta model} 
\label{sec:delta_mode_growth}

In the simplest scenario, we take the non-tachyonic phase to be infinitely short, so the peak in $\omega_k^2$ is described by $f(t) = \Lambda \delta(t)$, that is,
\be\label{eq:omega2_delta}
	\omega_k^2 
	= -\Gamma_k^2 + \Lambda\sum_{j \in \mathbb{Z}} \delta(t - jT)\,,
\ee
where $\Lambda$ is positive and $j$ enumerates the periods. We will refer to this ansatz as the \emph{delta model}.
Between the peaks, $\omega_k^2 = - \Gamma_k^2$ is constant and, by the mode equation \eqref{eq:dphi_eom_flat}, the perturbations evolve as
\be \label{eq:deltaphi_delta}
    \delta\phi_k = A_{j} e^{\Gamma_k (t-j T) } + B_j e^{-\Gamma_k (t-j T)} \, .
\ee
Continuity of $\delta\phi_k$ together with the mode equation implies that the first derivative must jump as $\Delta \delta\dot{\phi}_k = -\Lambda \delta\phi_k$ after encountering each peak. Then the coefficients in Eq.~\eqref{eq:deltaphi_delta} are joined together as
\be \label{eq:A_B_matrix}
	\begin{pmatrix} A_{j+1} \\ B_{j+1} \end{pmatrix} 
=  G \begin{pmatrix} A_j \\ B_j \end{pmatrix} \,
\equiv   \begin{pmatrix}
    \left( 1- \frac{\Lambda}{2\Gamma_k} \right) e^{\Gamma_k T}   & - \frac{\Lambda}{2\Gamma_k} e^{\Gamma_k T} \\ 
	\frac{\Lambda}{2\Gamma_k} e^{-\Gamma_k T}      & \left( 1 + \frac{\Lambda}{2\Gamma_k} \right) e^{-\Gamma_k T}
    \end{pmatrix} \begin{pmatrix} A_j \\ B_j \end{pmatrix}. 
\ee 
Although $G$ is not constructed via the Wronskian matrix as in Eq.~\eqref{eq:G}, its eigenvalues correspond to solutions that change by a constant factor after each period, so it is similar to the monodromy matrix. From Eq.~\eqref{eq:mu_from_trG} we obtain the spectrum of growth rates
\be\label{eq:muk_delta}
    \mu_{k} 
    = \frac{1}{T}{\rm Re}\,{\rm acosh}\left(\frac{1}{2}\tr G \right) 
    = \frac{1}{T}{\rm Re}\,{\rm acosh}\left(\cosh(\Gamma_k T) - \frac{\Lambda}{2\Gamma_k}\sinh(\Gamma_k T)\right) \, .
\ee
Regardless of other details such as the inflaton's potential, perturbation growth must satisfy the condition~\eqref{eq:k_small_growth} at $k=0$. This fixes
\be \label{eq:Lambda_relation_delta}
    \Lambda 
    = 2\Gamma_0 \coth\left(\frac{\Gamma_0 T}{2} \right)
    \overset{\Gamma_0T \gg 1}{\approx} 2\Gamma_0 \, ,
\ee
so model depends on two dimensional parameters, $T$ and $\Gamma_0$. The number of parameters can be further reduced by appropriate rescalings to a single dimensionless parameter, \eg, $\Gamma_0 T$~\cite{Tomberg:2021bll}. Other dimensionless quantities such as $\lambda_k T$ can be expressed in terms of it.

\begin{figure}
\begin{center}
	\includegraphics[width=.83\linewidth]{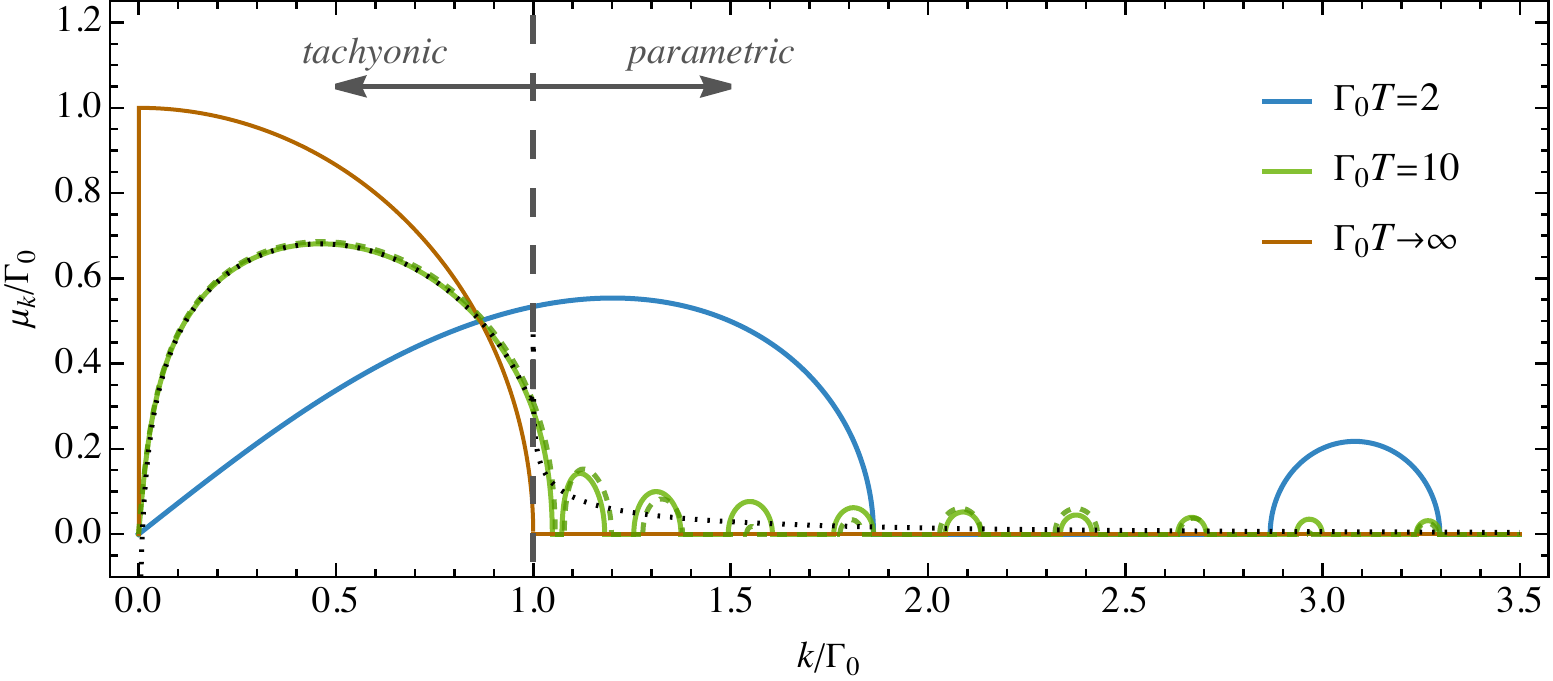}
\end{center}
\vspace{-4mm}
\caption{Delta model growth rate of modes for $\Gamma_0 T = 2$ (blue) and $\Gamma_0 T = 10$ (green) and $\Gamma_0 T \to \infty$ (orange). The gray dashed line shows the approximation \eqref{eq:lambda_approx} of the first peak for $\Gamma_0 T = 10$. To the right of the dashed vertical line, the instability stops being tachyonic. The dashed green line shows the box model with $T_1 = T/4$, $\Gamma_0 T = 10$ and $\tomega_0$ fixed by \eqref{eq:box_to_delta}.}
\label{fig:mu_k}
\end{figure}

The growth rates of different modes for different values of $\Gamma_0 T$ are shown in Fig.~\ref{fig:mu_k}. We observe a band structure in which unstable bands ($\mu_k > 0$) alternate with stable ones ($\mu_k = 0$). The tachyonic instability appears in the first and dominant band that begins at $k=0$. In the other bands, $k > \Gamma_0$ and thus $\omega_k^2 > 0$ at all times, so the instability is not tachyonic but due to parametric resonance. Since $\Gamma_k$ is imaginary there, the hyperbolic functions inside the brackets of Eq.~\eqref{eq:muk_delta} become trigonometric, causing $\cosh(\Gamma_k T)$ to oscillate and creating unstable resonance bands near the peaks where the subdominant $\sinh$ term lifts the absolute value of the sum above one. Eq.~\eqref{eq:muk_delta} implies that each band is terminated at $\Gamma_k T = j\pi i$, with $j$ a positive integer. Therefore, the first band lies in the range $0 < k < \sqrt{\Gamma_0^2 + \pi^2/T^2}$. Although there are infinitely many unstable bands at higher $k$, their growth rate is lower than that of the leading peaks, and they can often be neglected in practice. We discuss the UV behaviour in detail in section~\ref{sec:UV}.

The higher $k$ instability bands also become weaker with respect to the tachyonic band when $\Gamma_0$ increases. In the limit $\Gamma_0 T \to \infty$, shown by the orange curve in Fig.~\ref{fig:mu_k}, they disappear altogether, and the growth rate approaches the fully tachyonic $\mu_k \to \Gamma_k$, as expected. This limit corresponds to fast fragmentation due to a constant tachyonic mass term appearing like, \eg, in hilltop inflation~\cite{Felder:2000hj}. In the opposite limit $\Gamma_0 T \ll 1$, the first band gets broader but has a weaker instability.

In the mostly tachyonic case $\Gamma_0 T \gg 1$, which describes well the models we wish to study, the shape of the first peak is nicely approximated by
\be \label{eq:lambda_approx}
    \mufull_k \approx \Gamma_k + T^{-1} \ln\left[ 1 - \frac{\Gamma_0}{\Gamma_k} 
    \right],
\ee
as can be seen from Fig.~\eqref{fig:mu_k}. The fastest growing mode lies at
\be \label{eq:k_pk_approx}
    k_{\pk} \approx \sqrt{2\Gamma_0/T}\left[ 1 + \mathcal{O}\left(\Gamma_0 T\right)^{-1} \right]
\ee
and grows with the rate
\be \label{eq:mu_pk_approx}
    \mu_{\pk} \approx \Gamma_0 - T^{-1}(1 + \ln(\Gamma_0 T)) 
    + \mathcal{O}\left(\Gamma_0^{-1} T^{-2}\right).
\ee
These approximations agree with our previous observations. In particular, we can use them to compare a pure tachyonic instability, that is, a constant but negative mass term $-\Gamma_0^2$, with an oscillating mass with a strong tachyonic region. The purely tachyonic limit $\Gamma_0 T \to \infty$ can be obtained by sending the period $T$ to infinity while keeping $\Gamma_0$ constant, giving $\mu_\pk \to \Gamma_0$, $k_\pk \to 0$. In comparison, once we start decreasing $T$, the first peak's position shifts towards larger $k$, and its height decreases. That is, the instability bands become wider and flatter, as shown in Fig.~\ref{fig:mu_k}.

\subsubsection{Generating potential}
\label{sec:delta_background}

Above, we postulated an idealized time-dependence \eqref{eq:omega2_delta} for $\omega_k^2$ that allowed us to study mode growth analytically. We can go one step further and ask whether we can construct a potential that induces the time dependence \eqref{eq:omega2_delta} of $\omega_k^2$. The answer is positive: \eqref{eq:omega2_delta} is generated with the potential 
\be\label{eq:U_delta}
    U = -\frac{\Gamma_0^2}{2} \left((|\phi|-\phi_2)^2 - \phi_2^2\right),
\ee
if one also imposes the condition \eqref{eq:Lambda_relation_delta}, \ie, that the $k=0$ mode is stable. To see this, first note that the effective mass corresponding to the potential \eqref{eq:U_delta} is $U'' = - \Gamma_0^2 + 2 \Gamma_0^2 \phi_2 \delta(\phi)$. As the background crosses the origin periodically, we have 
\be
    \delta(\phib(t)) = |\dot\phib|^{-1} \sum_{j \in \mathbb{Z}} \delta(t - jT) \, ,
\ee
and comparing $U''$ to the ansatz~\eqref{eq:omega2_delta}, we have a match if $\Lambda = 2\Gamma_0^2 \phi_2/ |\dot\phib(0)|$.

To make the match explicit, we want to express all other quantities in terms of the parameters $\Gamma_0$ and $T$. As we work in the limit of negligible Hubble friction, the background's energy density $\rhob = U(\phi_\amp)$ is conserved and thus $|\dot{\phib}(\phib)| = \sqrt{2(\rhob - U(\phib))}$. Integrating this gives the half-period of $\phib$,
\be \label{eq:delta_T}
    T 
    = \frac{2}{\Gamma_0} {\rm acoth}\left(\frac{\Gamma_0\phi_2}{\sqrt{2\rhob}}\right) \, .
\ee
The velocity at the minimum is $|\dot{\phib}(0)| = \sqrt{2 \rhob}$. By combining everything, we find
\be \label{eq:Lambda_relation_delta_2}
    \Lambda = 2 \Gamma_0 \coth\left( \frac{T \Gamma_0}{2}\right), \,
\ee
which matches the condition \eqref{eq:Lambda_relation_delta} for the stability of the $k=0$ mode. We find that, for the delta model, the stability of the $k=0$ mode is equivalent to the existence of a potential from which the ansatz~\eqref{eq:omega2_delta} for $\omega_k^2$ can be generated. 

By relating the oscillation time $T$ and the amplitude, Eq.~\eqref{eq:delta_T} gives the final ingredient for determining background evolution. Nevertheless, by starting from $\omega_k^2$, it is not possible to fix the absolute scale of the field, but only the ratio $\phi_\amp/\phi_2$ can be fixed. Varying $\phi_2$ affects the height of the potential \eqref{eq:U_delta}, as can be seen in the bottom right panel of Fig.~\ref{fig:amp_3_comparisons}.

\subsection{Box model} 
\label{sec:box_model}

Generalizing the previous case, we can account for the duration of the non-tachyonic phase $T_1$. This leads us to model the peak in $\omega_k^2$ with the ansatz
\be\label{eq:f_box}
    f_{\rm box}(t) 
    = (\tomega_0^2 - \Gamma_0^2) \theta\left( T_1/2 - |t|\right) 
    - \tLambda \,\delta\left( |t|-T_1/2 \right)\, ,
\ee
where $\theta$ denotes the unit step function. In the tachyonic phase with length $T_2$, we now have $\omega_k^2 = k^2-\Gamma_0^2 = -\Gamma_k^2$ as before, and in the non-tachyonic phase with length $T_1$, $\omega_k^2 = k^2 + \tomega_0^2 \equiv \tomega_k^2$. The $\delta$-function is now located at the transition between the phases. It mimics the dip in $\omega_k^2$ before the onset of the tahyonic phase seen, for instance, in Figs.~\ref{fig:U_schematic} and~\ref{fig:amp_3_comparisons}. The stability of the $k=0$ mode imposes a condition that fixes $\tLambda$ in terms of the other 4 free parameters of the model: $T_1$, $T_2 \equiv T - T_1$, $\tomega_0$, and $\Gamma_0$.

The mode equation \eqref{eq:dphi_eom_flat} can be solved using standard methods for a piecewise constant $\omega_k^2$. The modes evolve as
\be\label{eq:deltaphi_box}
	\delta\phi_k = \left\{
\begin{array}{ll}
     A^{(1)}_j e^{i \tomega_k (t-j T) } + B^{(1)}_j e^{-i \tomega_k (t-j T)}, 
     &  \qquad  |t-j T| < T_1/2 \\
     A^{(2)}_j e^{\Gamma_k (t-j T) } + B^{(2)}_j e^{-\Gamma_k (t-j T)},
     &  \qquad T_1/2 \leq |t-j T| < T/2
\end{array}	\right. \, .
\ee
As in Eq.~\eqref{eq:A_B_matrix}, the monodromy matrix is given by the transformation connecting the $A^{(i)}_j$ coefficients. Appropriate matching, demanding the continuity of $\delta\phi_k$ and the discontinuity $\Delta \delta\dot{\phi}_k = \tLambda \delta\phi_k$ during transitions, then gives
\bea\label{eq:trG_box}
    \frac{1}{2}\tr G 
&    = \left[ \cos (T_1 \tomega_k ) + \frac{\tLambda}{\tomega_k} \sin (T_1 \tomega_k ) \right] \cosh (\Gamma_k  T_2)  \\
&    +   \left[ \frac{\Gamma_k^2-\tomega_k^2 + \tLambda^2}{2\Gamma_k \tomega_k } \sin (T_1 \tomega_k ) + \frac{\tLambda}{\Gamma_k} \cos (T_1 \tomega_k ) \right] \sinh (\Gamma_k T_2)
\eea
and, by Eqs.~\eqref{eq:mu_k},~\eqref{eq:mu_from_trG}, it gives the growth rates as $\mu_k = T^{-1}\Re \, {\rm acosh}(\tr G/2)$. The stability of the $k=0$ mode, \ie, $\tr G(k=0) = -2$, gives
\be\label{eq:Lambda_relation_box}
    \tLambda = \tomega_0 \tan\left(\frac{T_1 \tomega_0}{2}\right) - \Gamma_0 \coth\left(\frac{T_2 \Gamma_0}{2}\right)\, ,
\ee
eliminating one of the 5 parameters of the initial ansatz \eqref{eq:f_box}.

The band structure \eqref{eq:trG_box} of the box model is relatively similar to the one of the delta model \eqref{eq:muk_delta}. The most noticeable difference is the $k$-space modulation with period $k \approx \pi/T_1$ when $k\gtrsim \tomega_0$ that arises due the introduction of a new scale $T_1$. A comparison between the delta and the box models is shown in Fig.~\ref{fig:mu_k} by the solid and the dashed green lines, respectively. Although these lines almost overlap for the first three peaks, the fourth instability band is seen to be much weaker due to the modulating terms. Since the leading peak in $\mu_k$ tends to be similar in both models, the added complexity of the box model is most useful for obtaining a more accurate description of the subleading peaks.

\subsubsection{The $T_1 \to 0$ limit}

As a consistency check, we consider the limiting case of an instantaneous non-tachyonic phase, $T_1 \to 0$. In this process, we fix $\int f_{\rm box}(t)\,\td t = (\tomega_0^2 - \Gamma_0^2) T_1 - 2 \tLambda$ and allow for the possibility that $\tomega_0 \to \infty$. We find that this limit reproduces both the Floquet spectrum \eqref{eq:muk_delta} and the stability condition \eqref{eq:Lambda_relation_delta} of the delta-model with
\be\label{eq:box_to_delta}
    (\tomega_0^2 - \Gamma_0^2) T_1 - 2 \tLambda = \Lambda
\ee
or equivalently, with $\int f_{\rm box}(t)\,\td t = \int f_{\rm delta}(t)\,\td t$. 

Moreover, this matching permits us to compute the small $T_1$ corrections to the delta model: the trace of the monodromy matrix is
\be
    \frac{1}{2}\Tr G =
    \left(1 - \frac{T_1 \Lambda}{2}\right)\cosh(\Gamma_k T) - \frac{\Lambda}{2\Gamma_k}\left(1 - \frac{T_1 \Lambda}{4}\right)\sinh(\Gamma_k T) + \mathcal{O}(T_1^2)\, ,
\ee
while zero-mode stability gives
\be
    \Lambda 
    = 2\Gamma_0 \coth\left(\frac{\Gamma_0 T}{2} \right) - T_1 \Gamma_0^2 + \mathcal{O}(T_1^2).
\ee
Notice that these approximations hold as long as $T_1 \tilde\omega_k \approx T_1 k \ll 1$. Thus, it can fail for modes that oscillate with frequencies similar or larger to $1/T_1$, as these modes begin to probe the structure of the peak. In fact, this is a general limit of validity for these approximate models. As we will show in section \ref{sec:UV}, their growth rates $\mu_k$ decrease much slower at large $k$ than in scenarios based on smooth physical potentials.

\subsubsection{Generating potential}
\label{sec:box_background}

The potential that generates the box model \eqref{eq:f_box} can be constructed by glueing together quadratic functions,
\be\label{eq:U_box}
    U(\phi) 
    = \left\{
\begin{array}{lr}
    \frac{1}{2}\tomega_0^2 \phi^2, &                 \qquad |\phi| \leq \phi_1 \\
    -\frac{1}{2}\Gamma_0^2 (|\phi|-\phi_2)^2 + C, &  \qquad |\phi| > \phi_1 \\
\end{array}
\right. \, ,
\ee
where the constant $C = \frac{1}{2}m_1^2 \phi_1^2 + \frac{1}{2}m_2^2 (\phi_1-\phi_2)^2$ is fixed by demanding continuity at $|\phi| = \phi_1$. For the second derivative of the potential
\be
    U''(\phi) = \left(\Gamma_0^2 (\phi_1 - \phi_2) - \tomega_0^2 \phi_1\right) \delta(|\phi| - \phi_1) +
    \left\{
\begin{array}{l}
    \tomega_0^2, \qquad |\phi| \leq \phi_1 \\
    -\Gamma_0^2, \qquad |\phi| > \phi_1 \\
\end{array}
\right.
\ee
to give rise to \eqref{eq:f_box}, we must demand that
\be\label{eq:Lambda_relation_box2}
    \tLambda 
    = \frac{ \tomega_0^2 \phi_1 - \Gamma_0^2 (\phi_1 - \phi_2) }{|\dot\phib|(\phi_1)}
    = \tomega_0 \tan\left(\frac{T_1 \tomega_0}{2}\right) - \Gamma_0 \coth\left(\frac{T_2 \Gamma_0}{2}\right) \, ,
\ee
equivalent to \eqref{eq:Lambda_relation_box}, where, as with the delta model, we used $|\dot\phib (\phib)| = \sqrt{2(\rhob - U(\phib))}$ and integrated $|\dot\phib (\phib)|$ to get the durations of each epoch,
\be \label{eq:box_T1_T2}
    T_1 
    = \frac{2}{\tomega_0} \atan\left(\frac{\tomega_0\,\phi_1}{|\dot\phib|(\phi_1)}\right),
    \qquad
    T_2 
   = \frac{2}{\Gamma_0} {\rm acoth}\left(\frac{\Gamma_0 (\phi_2 -\phi_1)}{|\dot\phib|(\phi_1)}\right)\, .
\ee
The first derivative of the potential $U'$ is continuous if and only if $\tLambda = 0$. Again we find that imposing the condition \eqref{eq:Lambda_relation_box} allows us to consistently construct a potential from which the ansatz \eqref{eq:f_box} follows. Analogously to the delta model, \eqref{eq:box_T1_T2} fixes the ratios $\phi_\amp/\phi_2$ and $\phi_\amp/\phi_1$, completing the matching between $\omega_k^2$ and the background evolution. The overall scale of the field, and thus the height of the potential, is again left free.

\section{Modelling the band structure}
\label{sec:band_structure}

We aim to use the simplified models derived in the previous section to provide a detailed description of tachyonic preheating that is applicable to realistic scenarios on both the qualitative and the quantitative level.

As an example, we will use the simplified setups to model the instability bands of preheating with the potential
\be \label{eq:tanh2_pot}
    U(\phi) = U_0 \tanh^2 (\phi/\phi_0) \, ,
\ee
to which we refer to as the $\tanh^2$ potential. This potential represents the more general class of potentials with an exponentially flat plateau, that is, $U \sim U_0 (1 - A \exp(-2\phi/\phi_0))$, when $\phi \gtrsim \phi_0$. Here $A$ and $\phi_0$ are constants.
All such potentials lead to similar preheating phases~\cite{Tomberg:2021bll}. Plateau potentials can arise, \eg, from scalar field models of Palatini gravity with non-minimal gravitational couplings~\cite{Enckell:2018hmo, Antoniadis:2018ywb, Tenkanen:2020dge}, or string theory inspired alpha-attractor models~\cite{Ferrara:2013rsa, Kallosh:2013yoa, Carrasco:2015rva, Galante:2014ifa, Kallosh:2013hoa}. The perturbations have a tachyonic instability when $\phi \gtrsim 0.67\phi_0$ with a corresponding mass scale 
\be \label{eq:mth}
    \mth^2 \equiv U_0/\phi_0^2 \, .
\ee 
Applying the condition \eqref{eq:tach_cond} for an effective tachyonic instability, we find that $\Delta \phi_1 \sim \Delta \phi_2 \sim \phi_0$, $\Delta U_1 \sim \Delta U_2 \sim U_0$. As a result, we find that efficient tachyonic preheating is possible when 
\be
    \phi_0 \ll 1 \, ,
\ee
consistent with the numerical analysis of~\cite{Tomberg:2021bll} where the more precise limit $\phi_0 \lesssim 0.01$ was given. In this case, the potential \eqref{eq:tanh2_pot} predicts the scalar power spectrum $A_s \approx N^2 \mth^2/(3 \pi^2)$, the spectral index $n_s \approx 1 - 2/N$, and the tensor-to-scalar ratio $r \approx 2\phi_0^2/N^2$, when there are $N$ e-folds of inflaton left after the CMB pivot scale exits the cosmic horizon. With $N \approx 50$ and $\mth \approx 5\times10^{-6}$, the model fits the latest observational constraints~\cite{Planck:2018jri,BICEP:2021xfz}. In such plateau models, the tachyonicity requirement $\phi_0 \lesssim 0.01$ together with the fixed $\mth$ sets the limits $r \lesssim 10^{-7}$ and $U_0^{1/4} \lesssim 10^{14}$ GeV, giving an upper limit for the reheating temperature.

\subsection{Leading peaks}
\label{sec:leading_peaks}

\begin{figure}
    \centering
    \hspace{-1mm}
    \includegraphics[scale=0.85]{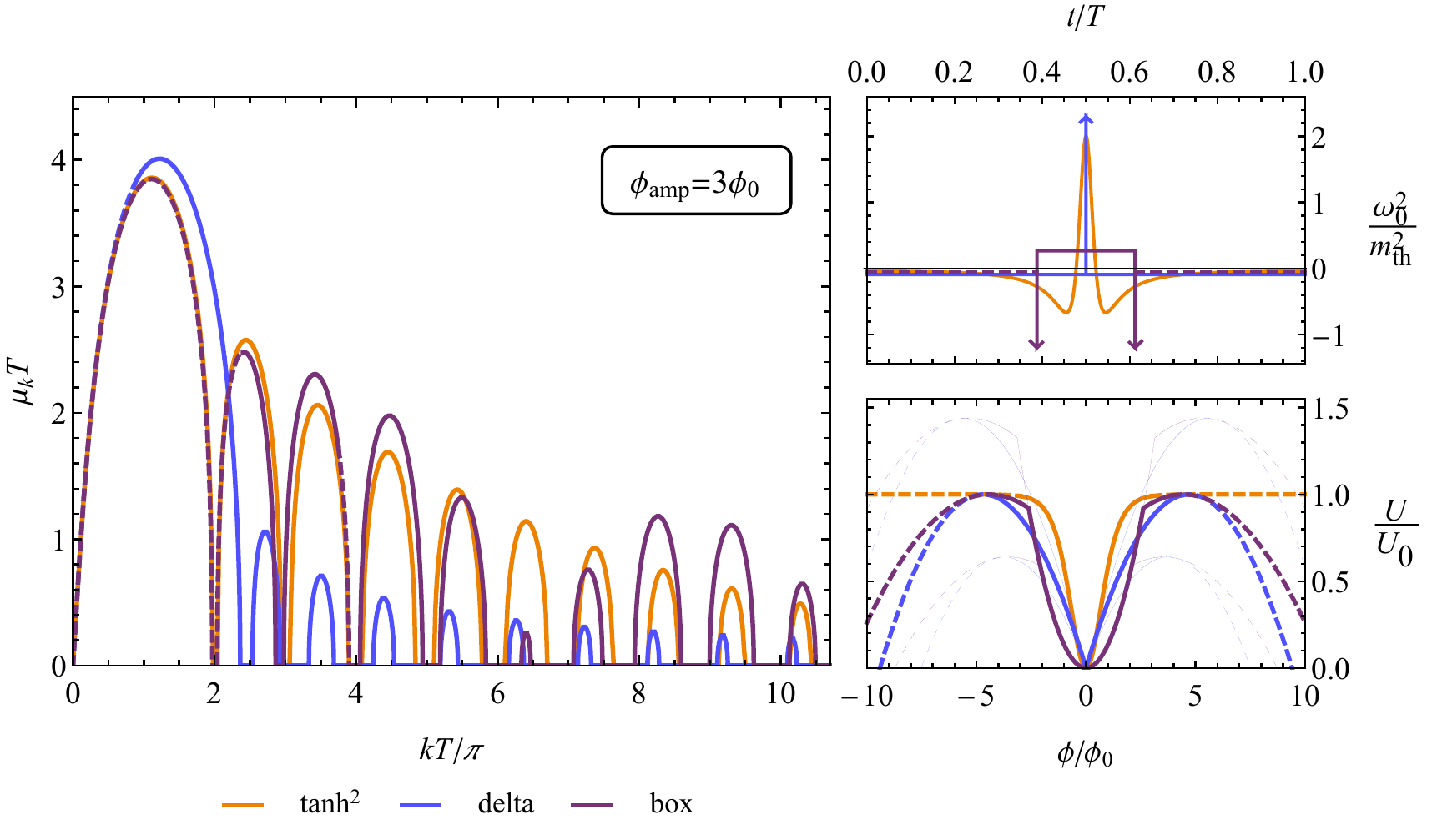}
    \caption{Comparison of the delta (blue) and box model (purple) fitted to the $\tanh^2$ potential \eqref{eq:tanh2_pot} (orange) with amplitude $\phi_\amp = 3\phi_0$. \emph{Left panel:} The corresponding growth rates $\mu_k$. \emph{Upper right panel:} The time-dependence of $\omega^2_0$. The arrows represent delta functions. \emph{Lower right panel:} The generating potential. The solid lines indicate the region where the field oscillates. Parameters for the delta model fit: $\Gamma_0 T = 6.72$ (giving $\phi_\amp = 0.93 \, \phi_2$), and for the box model fit: $\Gamma_0 T = 4.88$, $\tilde{\omega}_0 T = 11.61$, $T_1 = 0.22 \, T$, $T_2 = 0.78 \, T$ (giving $\phi_\amp = 0.88 \, \phi_2$ and $\phi_2 = 1.70 \, \phi_1$). The $\tanh^2$ potential uses $\phi_0 = 1.29\times 10^{-3}$, $U_0 = 4.15\times 10^{-17}$ (corresponding to an initial oscillation amplitude $\phi_\amp = 3\phi_0$~\cite{Tomberg:2021bll}, $\mth = 5\times 10^{-6}$, and $T=4.47\times 10^6$). Fixing $T$ determines the generating potentials up to a total scaling of $\phi$. In the lower right panel, three alternatives are depicted: one where the height of the potential is equal that of the $\tanh^2$ case ($\phi_2 = 6.06\times 10^{-3}$ for the delta and $\phi_2 = 5.72\times 10^{-3}$ for the box), and two where the $\phi_2$ is either $20\%$ higher or lower.
    }
    \label{fig:amp_3_comparisons}
\end{figure}

The speed of fragmentation dynamics, as well as the initial spectrum of perturbations, is determined mainly by the highest peaks in $\mu_k$, which lie at low $k$.\footnote{In the tachyonic cases considered here, the first $\mu_k$ peak (with the lowest $k$) is always the highest one. However, this is not generally true, \eg in the low amplitude non-tachyonic limits of the $\tanh^n$ models studied in \cite{Tomberg:2021bll} and for the orthogonal fields considered in section \ref{sec:multi_field} one of the subsequent peaks rises above the first.} In this section, we will compare the leading instability bands of the $\tanh^2$ model \eqref{eq:tanh2_pot} to the simplified scenarios. To this purpose, we fix the parameters of the simplified models as follows: 
\begin{enumerate}
    \item We demand that all models in the comparison have the same period $T$. For the $\tanh^2$ model, the period and the amplitude are related by $\mth T = \pi {\rm cosh}(\phi_{\rm amp}/\phi_0)$. Matching the period $T$ leads to similar $k$-space periodicity of the instability bands. In fact, this agreement between the periods in $\mu_{k}$ becomes exact in the limit $k\to \infty$.
    
    \item We fix the initial slope $\partial_k \mu_{k=0}$ of the growth exponents. For the $\tanh^2$ model, it can be obtained analytically from Eq.~\eqref{eq:k_small_growth} as
    \be \label{eq:muk_ini_slope}
        \partial_k \mu_{k=0} = \sqrt{ \sqrt{2}\mth T/\pi - 1} \,
    \ee
    even though the rest of the $\mu_k$ must be computed numerically. For the delta and the box models, the initial slope is easily obtained from Eqs.~\eqref{eq:muk_delta} and \eqref{eq:trG_box}, respectively.
    
\end{enumerate}
This procedure completely fixes the parameters of the delta model while it leaves two of the parameters 
free for the box model. The remaining parameters of the box model were fitted by eye to match the leading peaks of the box and $\tanh^2$ models as closely as possible.

An example is shown in Fig.~\ref{fig:amp_3_comparisons}, where we compare the $\mu_k$ spectrum, $\omega_k^2$, and the generating potentials of the three cases. The $\tanh^2$ spectrum is computed numerically. 
The delta model reproduces the leading first peak reasonably well, but the following secondary peaks are lower and narrower than the $\tanh^2$ counterparts.
The additional freedom of the box model allows for a practically perfect match for the leading peak and yields a better fit for the secondaries, but we see additional modulation in their heights, arising from the two competing time scales $T_1$ and $T_2$. Such modulation is absent in the smooth $\tanh^2$ case. Also, the models differ considerably in their large-$k$ behaviour, as is discussed in the following subsection. In the best fit box model, the non-tachyonic regime is relatively wide compared to  the $\tanh^2$ model, $T_1=0.28 \, T_2$. This is necessary to differentiate the box from the delta and to make the secondary peaks higher and wider.

\begin{figure}
    \centering
    \includegraphics{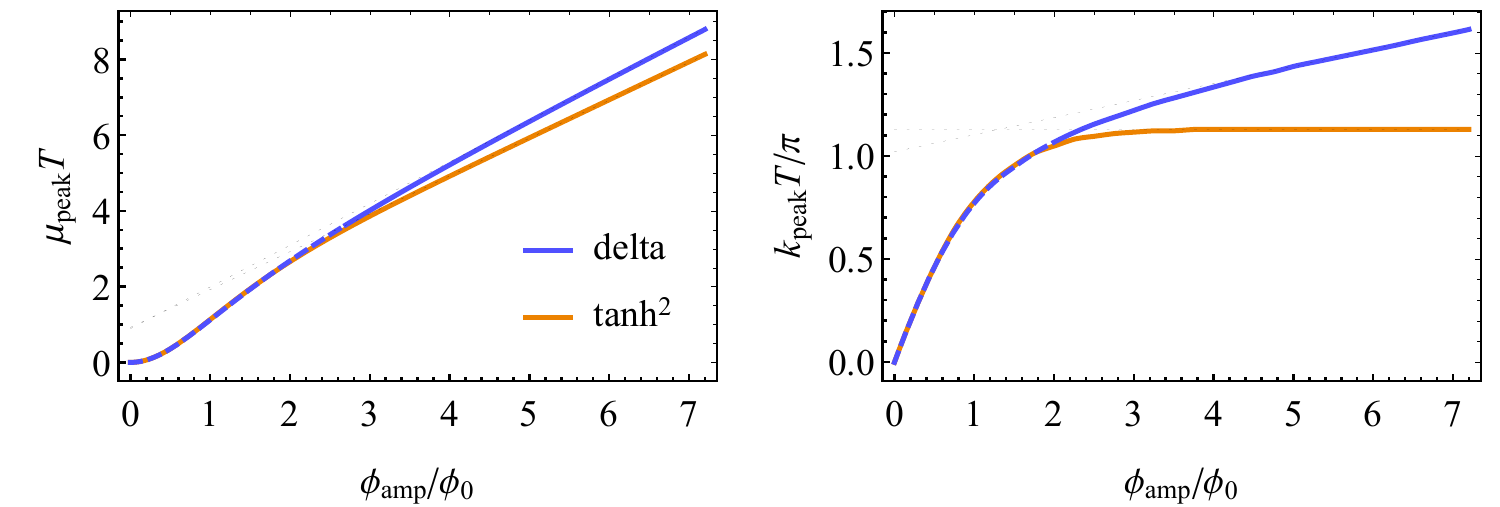}
    \caption{The height and location of the leading first $\mu_k$ peak for $\tanh^2$ model and for the corresponding delta fit, as explained in the text. The asymptotic behaviour (dotted black lines) for the $\tanh^2$ potential is $\mu_\text{peak} T \approx 0.92 + 1.00 \, \phi_\amp/\phi_0$, $k_\text{peak}T/\pi \approx 1.13$, and for the delta model, $\mu_\text{peak} T \approx 0.89 + 1.10 \, \phi_\amp/\phi_0$, $k_\text{peak}T/\pi \approx 1.02 + 0.083 \, \phi_\amp/\phi_0$. For comparison, the value of $T$ varies from $2.22 \, \mth^{-1}$ to $1218.05 \, \mth^{-1}$ as $\phi_\amp/\phi_0$ changes from $0$ to $7$.}
    \label{fig:peak_mu_k_comparison}
\end{figure}

In the $\tanh^2$ model, preheating dynamics is regulated almost entirely by the first peak when $10^{-4} \lesssim \phi_0 \lesssim 10^{-2}$~\cite{Tomberg:2021bll}. In Fig.~\ref{fig:peak_mu_k_comparison}, we have compared the predictions for this peak for the $\tanh^2$ model and the delta fit over a wide range of $\phi_\mathrm{amp}$ values. The analytical delta model agrees remarkably well with the numerical results of the $\tanh^2$ model. This implies that the first peak is largely determined by the behaviour of $\omega_k^2$ on the tachyonic plateau, which the delta model captures well. The delta model is completely analytic and relies mainly on Eq.~\eqref{eq:muk_ini_slope} for fixing $\Gamma_0$ from the slope of $\mu_k$ at $k=0$. This method provides a robust way for estimating the quantities relevant for tachyonic preheating with minimal computational effort. The box model does not provide such simple fits because of its additional free parameters. 

\begin{figure}
    \centering
    \includegraphics{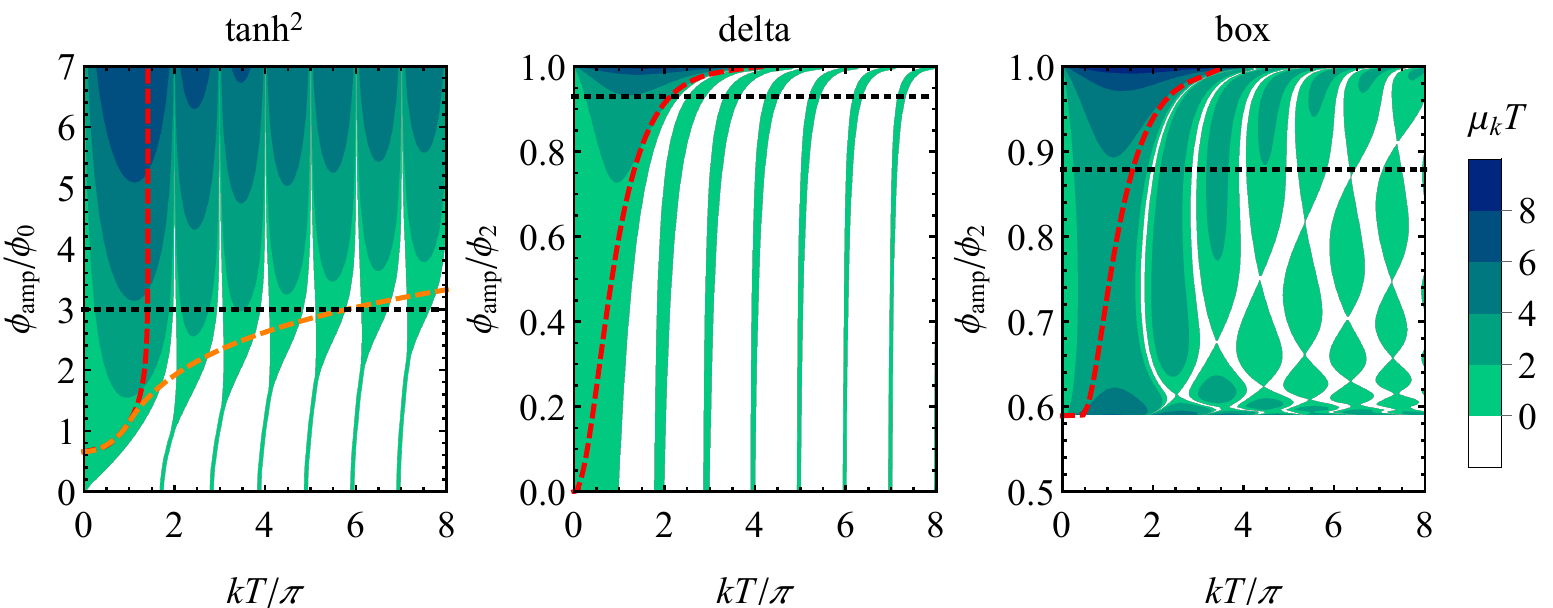}
    \caption{Floquet charts for the $\tanh^2$ potential \emph{(left panel)}, the delta model \emph{(middle panel)} and the box model \emph{(right panel)}. The red dashed curve indicates tachyonicity on the plateau: above it, $k^2 + U''(\phi_\amp) < 0$. For the $\tanh^2$ potential, the orange curve indicates overall tachyonicity: above it $\omega_k^2 < 0$ at the edge of the plateau where $U''$ is most negative. Notice the broadening of the secondary instability bands once a tachyonic contribution is present. The corresponding potentials are shown in Fig.~\ref{fig:amp_3_comparisons} and the black dotted line indicates the amplitudes used there.}
    \label{fig:floquet}
\end{figure}

Fig.~\ref{fig:floquet} shows the Floquet charts for the potentials of Fig.~\ref{fig:amp_3_comparisons} for all three models.\footnote{Notice a difference in methodology between Figs.~\ref{fig:peak_mu_k_comparison} and \ref{fig:floquet}: in \ref{fig:peak_mu_k_comparison}, we fit the delta model to the $\tanh^2$ model separately for each $\phi_\amp$ and plot the best-fit results, while in \ref{fig:floquet}, the fit is only done at one amplitude, the delta and box potentials are inferred from this, and then $\phi_\amp$ is varied within the fixed potential. The delta and box Floquet charts do not aim to match the $\tanh^2$ chart beyond the marked horizontal fit line.} We see that, in all models, only the first peak is fully tachyonic. The secondary peaks always have $\omega_k^2 > 0$ on the plateau at $\bar{\phi}=\phi_\amp$. The secondary $\tanh^2$ bands get wider as $\phi_\amp$ grows, consistently with the results of \cite{Tomberg:2021bll}, while the delta bands are always narrow, and the box bands exhibit modulation that varies with $\phi_\amp$. The strength of the secondary bands is tied to the properties of the dip in $\omega_k^2$ between the plateau and the non-tachyonic region. Due to this dip, the modes in these bands can still be partly tachyonic in the $\tanh^2$ model, leading to stronger peaks. The number of partly tachyonic peaks in the limit $\phi_\amp \gg \phi_0$ is 
\be \label{eq:Nth}
    N_{\rm th} 
    \approx \frac{1}{2\sqrt{3}}e^{\phi_\amp/\phi_0} \, ,
\ee
where we used the estimate $n \approx kT/\pi$ of the number of peaks below $k$ 
and that $k^2 < -U''_\mathrm{min} = 2/3$ at these peaks. For instance, when $\phi_\amp/\phi_0 \lesssim 2$, only the first peak is tachyonic, but for $\phi_\amp=5\phi_0$, over $100$ peaks are partly tachyonic. This is reflected in the Floquet chart in Fig.~\ref{fig:floquet}. As was noted in \cite{Tomberg:2021bll}, in the limit of high $\phi_\amp$, the secondary peaks play a major role in preheating and cannot be neglected. In comparison, there is no dip in the delta model, so the peaks are narrower and weaker. In the box model, the negative delta functions mimic the dip and produce stronger instability bands. Nevertheless, the match to the $\tanh^2$ model is not perfect. The detailed shape of the dip is essential for accurate modelling of the secondary peaks.

When $\phi_\amp$ is increased, the growth rate of modes within a single period, \ie, $\mu_k T$ gets larger. For the delta and box models, $\phi_\amp$ is limited from above by $\phi_2$, where the potential has a local maximum. In particular, the half-period $T$ diverges as $\phi_\amp \to \phi_2$. In the opposite limit $\phi_\amp \to 0$, the instability bands become increasingly narrower and weaker for the $\tanh^2$ potential. Beyond the leading band, the delta model is qualitatively similar. In the delta model, in the $T \ll 1/\Gamma_0$ limit we find a first band at $k \in (0, \pi/T)$ with height $\mu_k T = 0.79$. Curiously, the instability of the box model first increases as $\phi_\amp \to \phi_1 = 0.59 \, \phi_2$ because $\tilde{\Lambda}$ from \eqref{eq:Lambda_relation_box2} increases as $|\dot{\bar{\phi}}|(\phi_1)$ decreases, but then shuts off completely for $\phi_\amp < \phi_1$, since the potential is quadratic.

\subsection{UV instability bands}
\label{sec:UV}

Having discussed the first instability bands, let us turn our attention to the instability bands at large $k$. First, since the large $k$ modes behave almost adiabatically with $\dot{\omega}_k/\omega_k^2 \ll 1$, we have
\be\label{eq:G_adiab}
    \frac{1}{2} \tr G \sim \cos(\alpha_k(T)), \qquad 
    \alpha_k(t) \equiv \int_0^t \td t' \, \omega_k(t') \approx k t \, ,
\ee
that is, it oscillates between $-1$ and $1$. This result is derived by considering the independent solutions $u_{\pm}(t) \propto e^{ \pm i \alpha(t)}/\sqrt{\omega_k}$ to the mode equation and then applying Eq.~\eqref{eq:G} to compute the monodromy matrix. Since, by Eq.~\eqref{eq:mu_from_trG}, instability bands $\mu_k>0$ are possible only when $|\tr G/2|>1$, the adiabatic limit is stable, as expected. In other words, $\mu_k \to 0$ as $k\to\infty$.

According to the general theory of the Hill equation~\cite{1968ZaMM...48R.138R}, there is an infinite number of instability bands and $|\tr G|-2$ must decrease as $k^{-2}$ or faster when $k\to \infty$, implying that $\mu_k$ should diminish at least as fast as $k^{-1}$.\footnote{This follows from $\mu_k T \sim {\rm acosh}|\tr G/2| \sim \sqrt{|\tr G|-2}$ as $|\tr G|\sim 2$.} However, the $\mu_k \sim k^{-1}$ scaling in the UV would be too slow to be physical as it would lead to a catastrophic growth of the energy density of UV perturbations. To demonstrate this, consider initial growth of perturbations sourced by vacuum fluctuations, in which case the energy density grows as~\cite{Tomberg:2021bll}
\be \label{eq:j}
    \delta \dot \rho \approx j
    \equiv  \int \frac{\td^3 k}{(2\pi)^3} 2 \mu_k \delta\rho^{\rm vac}_{k}\, ,
\ee
where $\delta\rho^{\rm vac}_{k} = k/2$ is the vacuum energy density. This integral must be finite in well-behaved models. Otherwise, the linear description cannot be valid up to arbitrarily high $k$, and there must exist a cut-off scale $k_\mathrm{cut}$ regulating the growth of perturbations.\footnote{This implies a breakdown of the perturbative vacuum at high $k$, though this could be cured by non-linear effects.} Approximating the $n$-th peak in the spectrum of growth rates as 
\be\label{eq:muk_UV_peak}
    \mu_k \approx \mu_{\pk,n}\left(1 - 4\frac{(k -k_n )^2}{\Delta_{\pk,n}^2}\right),
\ee
where $\mu_{\pk,n}$ $k_n$, and $\Delta_{\pk,n}$ give the height, position and width of the peak, respectively, we can approximate
\be \label{eq:j_sum}
    j \sim \frac{1}{3\pi^2}\sum_n \mu_{\pk,n} \Delta_{\pk,n} k_n^3 \, .
\ee
As implied from extrema of $\tr G$ in Eq.~\eqref{eq:G_adiab}, it approximately holds that $k_n \propto n$. Thus, for the sum to converge, $\mu_{\pk,n} \Delta_{\pk,n}$ must die away faster than $k^{-4}$. In the following, we will show that this requirement is not satisfied by the simplified models. Therefore, they must be equipped with a UV cut-off on $\mu_{k}$. In practice, this is a benign issue, as the scales at which the problematic UV behaviour shows up are well separated from the scale of the tachyonic instability. Thus, since the fragmentation dynamics is governed by the first instability bands, imposing a UV cut-off has a negligible impact on the applicability of the simplified models beyond curing the far-UV divergence.

Furthermore, although Hubble friction can be mostly neglected for the leading bands, expansion may soften the UV instability by causing the growing high-$k$ modes to redshift out of the increasingly narrow instability bands. For any amount of expansion, modes with a sufficiently high $k$ will not stay on a given resonance band for even within a single oscillation of the background field --  periodicity of the mode equation cannot be assumed for such modes and the Floquet analysis breaks down. Particles may still be produced at high $k$ in the presence of a strong feature such as the delta peaks, so UV divergences may still appear, but they should be analysed using different techniques. Since the $k$ value where such effects kick in varies from case to case, we ignore this issue for the rest of the section. We proceed to study the UV behaviour of our models in the flat space limit, believing it to be of theoretical interest, if not directly relevant for the preheating dynamics.

In the UV, it is the small deviations from perfect adiabaticity that cause instability. As these deviations approach zero when $k\to\infty$, the maxima of $\mu_k$ must lie close to the extrema of $\cos(\alpha_k(T))$ from \eqref{eq:G_adiab}, that is, at
\be \label{eq:peak_positions}
    \alpha_{k_n}(T) = \pi n + \eps_n \, , \quad n \in \mathbb{Z}_+, \qquad n \gg 1 \, ,
\ee
where $\epsilon_n$ denotes small deviations from the adiabatic expectation. Indeed, the peaks get increasingly narrower so the difference is important especially in numerical searches, for which $\alpha_k(T) = \pi n$ also gives a considerably better initial guess than the simpler limiting case $kT=\pi n$. We look for the extrema of $\tr G/2$ of the form \eqref{eq:peak_positions} by considering the leading terms in the limits $n\gg 1$ and $\eps_n \ll 1$ and then maximizing with respect to $\eps_n$. A general expansion around the peaks then reads
\be\label{eq:UV_trG}
    \qty|\frac{1}{2}\tr G|
    \approx 1 - \frac{1}{2}\eps_n^2 + F_{0,n} + \eps_n F_{1,n}
    \, ,
\ee
where the first term arises from expanding the leading adiabatic contribution \eqref{eq:G_adiab} and the last two terms represent small non-adiabatic corrections to zeroth and first orders in $\eps_n$. At the leading order, we can ignore corrections of the form $\eps_n^2 F_{2,n}$, since we expect $F_{2,n} \ll 1/2$. The maximum of \eqref{eq:UV_trG} then lies at $\eps_n = F_{1,n}$, with $|\tr G/2| = 1 + F_{0,n} + F_{1,n}^2/2$. Expanding ${\rm acosh}(1+x) \approx\sqrt{2x}$ to obtain $\mu_k$ from \eqref{eq:mu_from_trG} and mapping this into the parameters of \eqref{eq:muk_UV_peak} we find that, at the leading order,
\be\label{eq:UV_peak_params}
    \mu_{\pk,n} = \frac{1}{T}\sqrt{2F_{0,n} + F_{1,n}^2} \, , \qquad
    \Delta_{\pk,n} = 2\mu_{\pk,n} \, , \qquad
    k_n = \pi n/T + F_{1,n} \, .
\ee
Thus the widths of the instability bands are diminished at the same pace as their heights. In particular, $j \propto \sum_n \mu_{\pk,n}^2 k_n^3$, so it is sufficient to require that $\mu_{\pk,n}$ decreases faster than $k^{-2}$ in the UV for $j$ to converge. 

Let us now give concrete examples of the UV instability bands in the simplified models and the $\tanh^2$ model.

\begin{figure}
    \centering
    \includegraphics[scale=0.7]{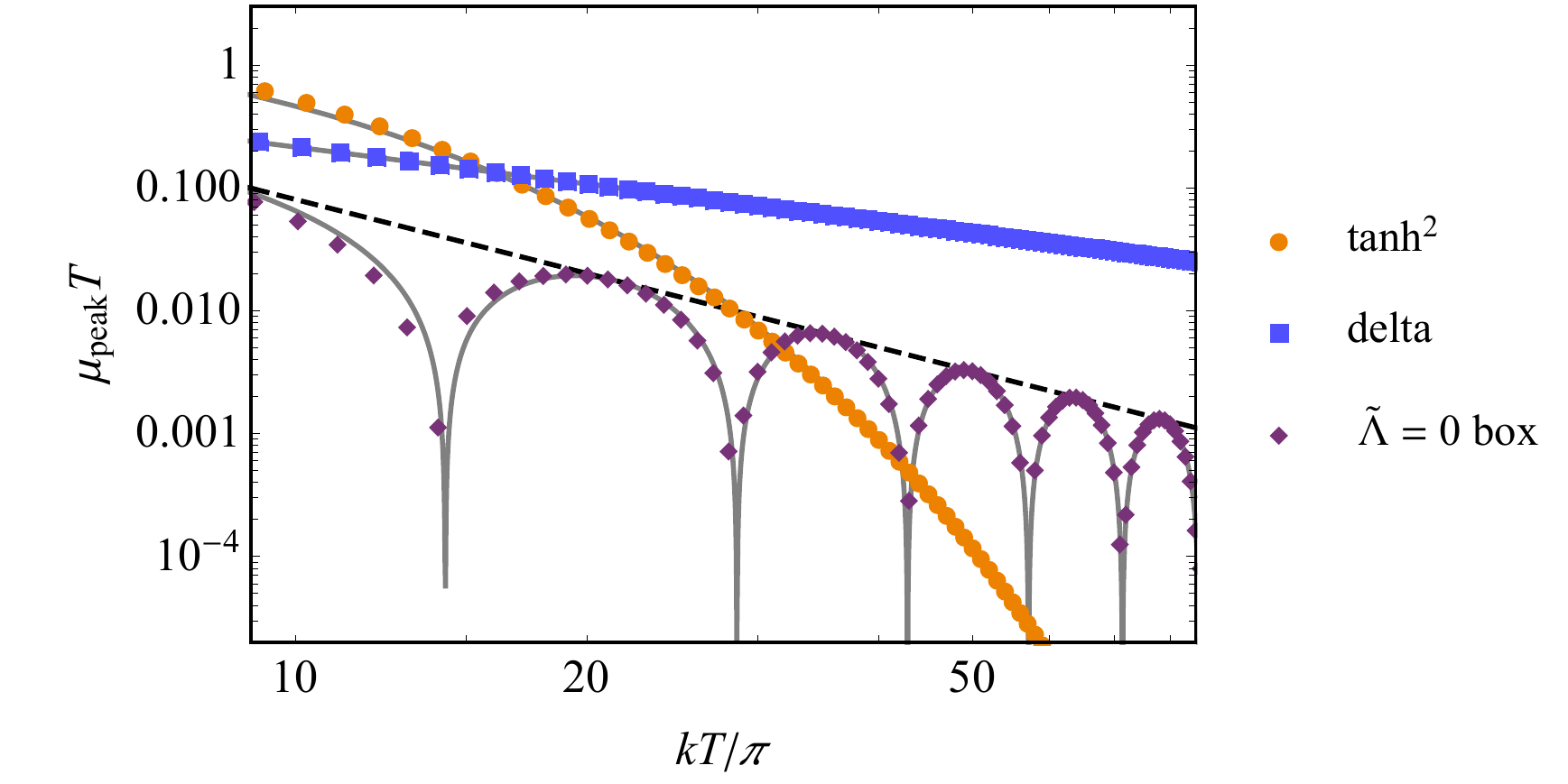}
    \caption{Positions of the peaks in $\mu_k$ in the large-$k$ tail. The example models are the same as those in Fig.~\ref{fig:amp_3_comparisons}, except for the $\tilde{\Lambda}=0$ box, which is obtained from the box model by adjusting $T_1/T$ so that $\tilde{\Lambda}=0$, achieved with $T_1 = 0.070 \, T$. The grey lines give fits to the data: for the $\tanh^2$ model, the numerical fit $\mu_\text{peak} T = 3.63\, e^{-1.47 k/\mth}$; for the delta model, the $k^{-1}$ scaling of \eqref{eq:UV_peak_delta}; and for the $\tilde{\Lambda}=0$ box model, the modulated $k^{-2}$ scaling of \eqref{eq:UV_peak_box}. The dashed black line gives the envelope of the modulated peaks for the box from the magnitude of \eqref{eq:UV_peak_box}. Notice the excellent matches between the peaks and the analytic estimates.}
    \label{fig:peaks}
\end{figure}

\subsubsection*{UV instability of the delta model}
\label{sec:UV_delta}

For the delta model, $\alpha(T) = -i\Gamma_k T$. Expanding the trace of the monodromy matrix ~\eqref{eq:muk_delta} around $\alpha(T) = \pi n$ then gives
\be \label{eq:approx_trG_delta}
    \qty|\frac{1}{2}\tr G| \approx 1 - \frac{1}{2}\eps_{n}^2 - \frac{\Lambda}{2k}\eps_{n} \, ,
\ee
where $k$ corresponds to the value at $\eps_{n}=0$. Comparing with Eq.~\eqref{eq:UV_trG}, we find $F_{0,n} = 0$ and $F_{1,n} = \Lambda/(2k)$, so that by Eq.~\eqref{eq:UV_peak_params} we have
\be \label{eq:UV_peak_delta}
    \mu_\pk = \frac{\Lambda}{2kT} + \mathcal{O}(k^{-2}) \, . 
\ee
This behaviour is shown in Fig.~\ref{fig:peaks}. The peak decreases too slowly, so the $j$ integral~\eqref{eq:j} diverges. The source of this slow damping is the delta function in the $\omega_k^2$ ansatz \eqref{eq:omega2_delta}, which can efficiently excite arbitrarily high $k$-bands through parametric resonance. In particular, the adiabaticity condition $\dot{\omega}_k/\omega_k^2 \ll 1$ is never completely satisfied due to this delta function.

\subsubsection*{Softening the UV instability with the box model}
\label{sec:UV_box}

For the box model, $\alpha(T) = \alpha_1 + \alpha_2$, where $\alpha_1 \equiv T_1 \tomega_k \sim k T_1$, $\alpha_2 \equiv T_2 \Gamma_k \sim k T_2$. At $k\gg1$, the trace of the monodromy matrix~\eqref{eq:trG_box} an be arranged as
\be\label{eq:approx_trG_box}
    \frac{1}{2}\tr G = \cos(\alpha(T)) 
    + \frac{\tLambda}{k}\sin(\alpha(T))
    - \frac{(\Gamma_0^2 + \tomega^2)^{2}}{8k^4}\sin(\alpha_1)\sin(\alpha_2) + \mathcal{O}(\Lambda^2,k^{-6}) \, , \\
\ee
where we included the leading term that survives it the $\tLambda\to 0$ limit. Replacing $\alpha(T)=\pi n +\eps_{n}$ and expanding in $\eps_{n}$ gives
\be
    \qty|\frac{1}{2}\tr G| \approx 1 - \frac{1}{2}\eps_{n}^2 + \frac{\tLambda}{k}\eps_{n} + \frac{(\Gamma_0^2 + \tomega^2)^{2}}{8k^4}\sin^2(T_1 k) \, ,
\ee
where the last two terms give us $F_{1,n}$ and $F_{0,n}$, respectively. If $\tLambda \neq 0$, then  $\mu_\pk \sim |\tLambda|/(kT)$ resembling the UV-behaviour of the delta model. However, when $\tLambda = 0$, we find
\be\label{eq:UV_peak_box}
    \mu_{\pk, \tLambda = 0} = \frac{\Gamma_0^2 + \tomega^2}{2k^2T}\qty|\sin(T_1 k)| + \mathcal{O}(k^{-3})\, , 
\ee
so the $j$ integral~\eqref{eq:j} diverges only logarithmically. Notice that the adiabaticity condition $\dot{\omega}_k/\omega_k^2 \ll 1$ is not satisfied at the transition between the tachyonic and non-tachyonic phases, \ie, at the boundary of the box, even if $\tLambda = 0$. Although this violation is weaker, it allows the parametric resonance to stay active at very high $k$ and provides a physical explanation for the divergence of $j$.

Another interesting feature is the modulation of $\mu_{\pk,\tLambda = 0}$ due to the secondary scale $T_1$, also shown in Fig.~\ref{fig:peaks}. Moreover, \eqref{eq:UV_peak_box} reproduces the delta model \eqref{eq:UV_peak_delta} in the $T_1 \to 0$ limit when we make the identification $T_1 \tomega^2 = \Lambda$. Notice that this matching differs slightly from Eq.~\eqref{eq:box_to_delta} since we set $\tLambda = 0$. In particular, the $k\to 0$ and $T_1 \to 0$ limits of $\mu_k$ do not commute.

\subsubsection*{The $\tanh^2$ potential and UV stability of smooth potentials}
\label{sec:UV_tanh2}

For the $\tanh^2$ model~\eqref{eq:tanh2_pot}, we must resolve the UV instability bands numerically. A numerical scan of amplitudes in the range $\phi_\amp \in (\phi_0,7\phi_0)$ reveals an exponential suppression\footnote{We expect this suppression to become effective in the fully non-tachyonic region which, for high $\phi_\amp$, means a high band number $n$, as discussed around Eq.~\eqref{eq:Nth}.}
\be\label{eq:UV_peak_tanh}
    \mu_\pk \approx \Delta_\pk/2 \sim e^{-1.5k/\mth} \, ,
\ee
with some percent level variation in the exponential slope lying within the numerical uncertainty of the computation. Consistent with Eq.~\eqref{eq:UV_peak_params}, the widths obey roughly $\Delta_\pk \approx 2\mu_\pk$. The case $\phi_\amp=3\phi_0$ is shown in Fig.~\ref{fig:peaks}. Due to the exponential damping of $\mu_k$, the $j$ integral~\eqref{eq:j} converges, and the model is free of instabilities in the UV. The damping scale $\mth$ can be understood intuitively -- it is the highest scale contributing to the time-dependence of $U''$. Effectively, $\mth$ acts as the cut-off scale of the model---modes with $k \gg \mth$ are increasingly adiabatic and stable.

Above we have observed that the degree of discontinuity in $\omega_k^2$ corresponds how fast $\mu_k$ decreases with $k$: if $\omega_k^2$ contains a $\delta$-function or a simple discontinuity, then the scalings $\mu_k \sim k^{-1}$ and $\mu_k \sim k^{-2}$ are obtained, respectively. A smooth $\omega_k^2$, on the other hand, leads to the exponential suppression of $\mu_k$. One can further show that a continuous $\omega_k^2$ leads to a decrease at least as fast as $\mu_k \sim k^{-3}$~\cite{1968ZaMM...48R.138R}, suggesting that in a model where the $n^\mathrm{th}$ derivative of $\omega_k^2$ is the first discontinuous one, the heights and widths of the peaks decrease as $k^{-n-2}$.\footnote{Although the UV-instability can be removed by using a piecewise continuous ansatz for $\omega_k^2$, the resulting UV-instability bands would still significantly differ from smooth physical models.} We further expect that $\mu_k$ is exponentially damped whenever $\omega_k^2$ is smooth, though we are not aware of a general proof.

\section{Multi-field scenarios}
\label{sec:multi_field}

\begin{figure}
    \centering
    \includegraphics[scale=0.85]{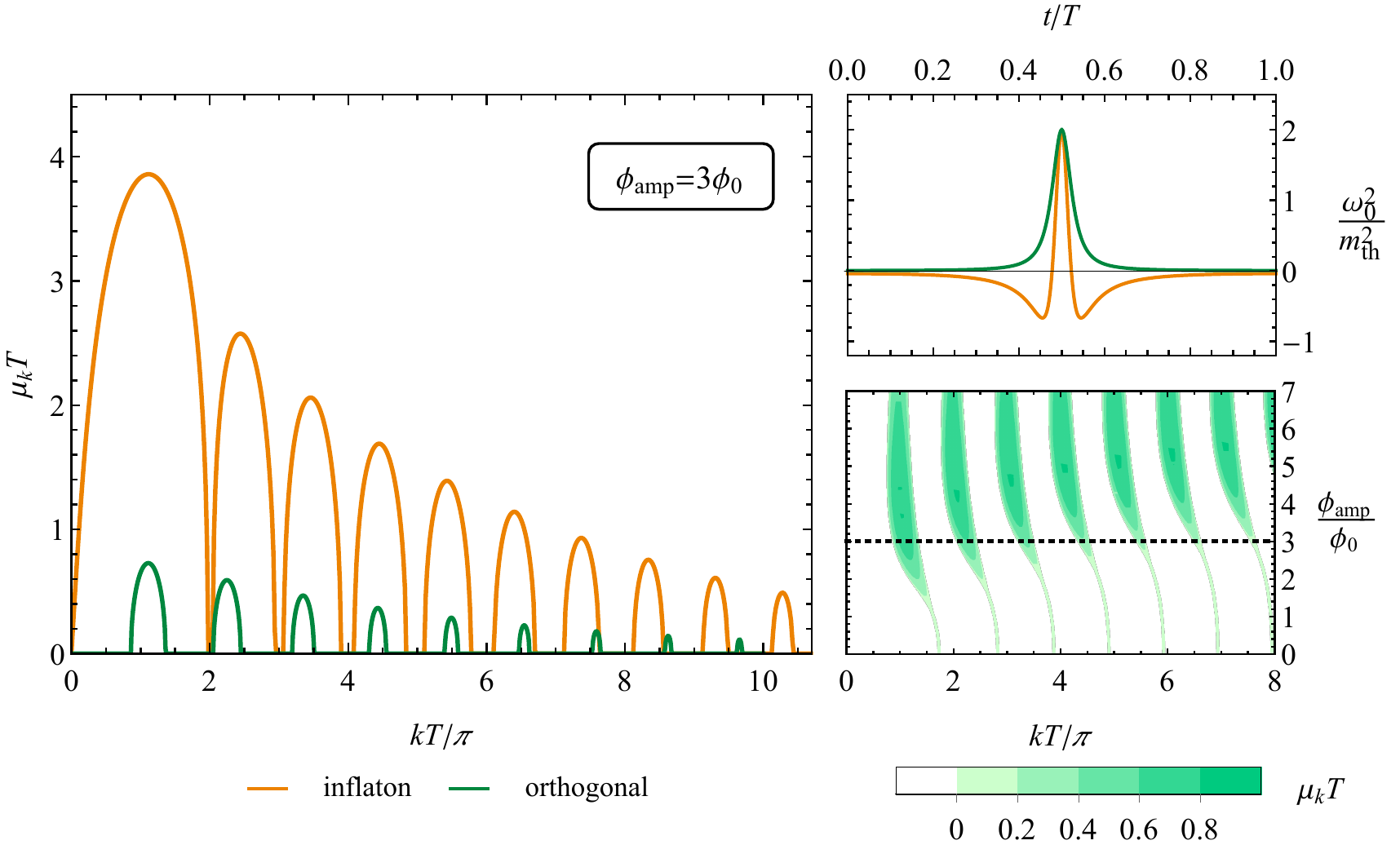}
    \caption{Comparison of mode growth of the inflation (orange) and orthogonal (green) perturbations for $O(N)$ inflation assuming the $\tanh^2$ potential.  \emph{Left panel:} An example of the $\mu_k$ spectra for $\phi_\amp = 3\phi_0$. \emph{Upper right panel:} The function $\omega_0^2$ for $\phi_\amp = 3\phi_0$. Note that the orthogonal component is never tachyonic. \emph{Lower right panel:} The Floquet chart for the orthogonal modes. The dashed horizontal line indicates $\phi_\amp = 3\phi_0$ depicted in the other panels.
    }
    \label{fig:secondary_field}
\end{figure}

Preheating can be more complicated when the inflaton is coupled to other fields since these too can be excited due to the oscillating background. It is also possible that the inflaton itself is not a simple real scalar. This section focuses on the latter possibility and considers an inflaton symmetric under a global $O(N)$ group as a simple multi-field example of tachyonic preheating.

As above, we assume fast preheating and neglect the effect of expansion. The $O(N)$ symmetry dictates that the potential $V(\phi)$ is a function of the field modulus $\phi \equiv \sqrt{\sum_i \phi^2_i}$ only and the background field equations read
\be
    \ddot{\phib}_i = - \hat\phi_i V',
\ee
where $\hat\phi_i \equiv \phib_i/\phib$, the prime denotes a derivative with respect to $\phi$, and a bar refers to the background field.\footnote{In contrast to the rest of the article, in this section $\bar{\phi}$ is the modulus of the background field and thus always positive.} We assume that the potential has a minimum at $\phi = 0$ and that it is a monotonously growing function of $\phi$, so that $V' > 0$. Expansion during inflation damps all angular motion in the field space, so the background field rolls down in a radial direction. We choose the basis so that the background is non-vanishing only along the $\phi_1$ component, $\bar \phi_i \propto \delta_{1i} = \hat \phi_i$, which we will refer to as the inflaton direction. The perturbations $\delta \phi \equiv \phi_i - \bar \phi_i$ evolve as
\be\label{eq:eom_ortho}
    \delta \ddot \phi_{i,k} + (m^2_{ij} + k^2 \delta_{ij}) \delta \phi_{j,k} = 0 \, , 
    \qquad \mbox{where} \quad
    m^2_{ij}
    \equiv \partial_{\phi_j,\phi_i}V \, .
\ee 
For $O(N)$ symmetric fields, the mass matrix
\be
    m^2_{ij} 
    = \hat \phi_i \hat \phi_i V'' + \left(\delta_{ij} - \hat \phi_i \hat \phi_i\right) V'/\phib
    = {\rm diag}(V'', V'/\phib, V'/\phib, \ldots)
\ee
is diagonal and the directions decouple from each other. Moreover, only the inflaton direction possesses a tachyonic instability through $V''<0$---for monotonous potentials, $V' > 0$, so the mass squared of orthogonal perturbations is positive. These directions are still subject to parametric instabilities. The corresponding mode equations are
\be\label{eq:mode_perp}
    \delta \ddot \phi_{\perp,k} + (V'/\phib + k^2) \delta \phi_{\perp,k} = 0 \, ,
\ee
leading to some growth rates $\mu_{\perp,k}$. The symbol $\perp$ refers to an orthogonal component with $i > 1$. These components are identical to each other. These modes have the following general features:
\begin{itemize}
    \item The $k = 0$ mode is not growing. This can be seen by noting that $V'/\phib = -\ddot{\phib}_1/\phib_1$ so that $\delta \phi_{\perp,0}(t) \propto \phib_1(t)$. Since the background is periodic in $\phib(t)$, then, analogously to the inflaton mode, its Floquet exponents must vanish~\cite{Tomberg:2021bll,1968ZaMM...48R.138R}.
    
    \item The first instability band of orthogonal modes must begin at some finite, non-zero $k$ (for a derivation, see appendix~\ref{app:SON}). This is to be contrasted with the perturbations in the tachyonic direction for which the first instability band begins already at $k=0$.\footnote{This holds when the oscillation period decreases with decreasing energy density (see Eq.~\eqref{eq:k_small_growth}).} As the band structure of both the inflaton and the orthogonal modes have a similar period in $k$-space, the first orthogonal modes must then be narrower. An example of these features is shown in Fig.~\ref{fig:secondary_field}.
    
    \item In the limit $k \to \infty$, the perturbations are adiabatic and thus the growth rates vanish. How fast $\mu_{\perp,k}$ approaches 0 depends on $V'/\phib$. If $V'/\phib$ is smooth, as for the $\tanh^2$ model, it's expected that $\mu_{\perp,k}$ is damped exponentially for large $k$ as was the case with the inflaton mode. By Eq.~\eqref{eq:G_adiab}, the corresponding monodromy matrix oscillates as $\cos(kT)$ in the $k \to \infty$ limit.
\end{itemize}

One could construct simplified models for the orthogonal modes, for instance, by relying on the generating potentials \eqref{eq:U_delta} and \eqref{eq:U_box}. However, the orthogonal mode equation \eqref{eq:eom_ortho} does not have a simple analytic solution for these potentials. Thus we found it more illuminating to rely on general analytic derivations and numerical methods. As an example, we worked out the case of the $\tanh^2$ potential~\eqref{eq:tanh2_pot} shown in Fig.~\ref{fig:secondary_field}.

In all, the orthogonal modes are subject to a narrow parametric resonance that is generally weaker than the broad tachyonic instability of inflaton modes. Thus, at least at the leading order, it is sufficient to consider only the inflaton modes when studying inflaton fragmentation in the linear regime.

\section{Discussion}
\label{sec:discussion}

In a large class of inflationary models with an exponentially flat plateau, including the $\tanh^2$ model, tachyonic preheating is efficient only when the tensor-to-scalar ratio is extremely small, $r<10^{-7}$ \cite{Tomberg:2021bll}. Preheating is also a potential source of gravitational waves (GWs)~\cite{Garcia-Bellido:2007nns, Garcia-Bellido:2007fiu, Dufaux:2008dn, Dufaux:2010cf, Antusch:2016con, Liu:2017hua, Adshead:2018doq, Amin:2018xfe, Adshead:2019lbr, Lozanov:2019ylm, Hiramatsu:2020obh, Cui:2021are}, and for tachyonic preheating in these potentials, the GW signal is expected to peak at frequency $\fgw=10^9$ Hz or above~\cite{Tomberg:2021bll}. Such predictions are useful for differentiating models of inflation from each other, but in this case, they also lie well beyond our current observational capabilities. It is then interesting to ask whether these predictions are generic to all inflationary models that transition into an epoch of rapid tachyonic preheating. We show that this is not the case, but instead, it is possible to construct inflationary models with fast tachyonic preheating that can produce virtually any $r$ and $\fgw$.

\paragraph{Inflationary observables.} 
Let us first review how the limits on the $\tanh^2$ preheating arise. As discussed in section~\ref{sec:band_structure}, the CMB scalar perturbation strength $A_s$ fixes the mass parameter $\mth^2=U_0/\phi_0^2 \approx 5 \times 10^{-6}$, which sets also the scale of tachyonicity during preheating, $|U''| \sim \mth^2$. Thus, in this case, the shape of the potential relates the scales relevant to CMB and preheating. Fast tachyonic preheating means that the background oscillation time $T$ during preheating must satisfy $T \ll H^{-1}$. Since $H\sim\sqrt{U_0}$ from the first Friedmann equation and $T \sim \mth^{-1}\phi_0^{-1/2}$~\cite{Tomberg:2021bll}, we get the condition $\phi_0 \ll 1$. A more precise analysis shows that $\phi_0 \lesssim 0.01$ is necessary and sufficient for exponentially flat potentials~\cite{Tomberg:2021bll}. As $\mth$ is fixed, we obtain that $U_0 \lesssim (5\times10^{14}\,\text{GeV})^4$ and $r \lesssim 8 \times 10^{-8}$. To put it briefly, an effective tachyonic preheating requires that the scale of inflation must be lower than the mass. On the other hand, if the mass is fixed as for the $\tanh^2$ potential, it may infer a small scale of inflation and thus a small $r$.

Let us consider a more general model, where the CMB and preheating scales are not intimately related. For instance, as a proof of concept, we may construct a potential by glueing good inflationary plateaus to the delta model potential. The shape of the plateau and the minimum can then be tuned independently. The condition \eqref{eq:tach_cond} for tachyonic preheating, $T \ll H^{-1}$, reads $\phi_2 \ll 1$, which simply means that \emph{efficient preheating requires sub-Planckian field excursions}. We should further impose that the mass scales are sub-Planckian, $\Gamma_0 \ll 1$, which gives the hierarchy
\be
    H \ll \phi_2 \ll 1 \, .
\ee
At this very general level of discussion, this is the strongest theoretical bound and leaves plenty of room for the current experimental bounds $H < 2\times 10^{-5}$~\cite{Planck:2018jri} to be satisfied\footnote{The bound on $H$ arises from $r < 0.036$~\cite{BICEP:2021xfz} in single field slow-roll inflation.}. The smaller the field scale $\phi_2$, the more difficult it will be for $r$ (or $H$) to saturate the observational bounds. Although this argument does not give an explicit physically viable inflationary model, it provides a proof of concept showing that tachyonic preheating can be consistent with any experimentally allowed set of inflationary parameters.

\paragraph{Induced gravitational waves.} 
Let us now consider GWs. Their frequency can be estimated as $\fgw \approx (z T)^{-1}$, where the redshift is $z \approx T_{\rm reh}/T_\mathrm{CMB}$. Here $T_{\rm reh}$ is the reheating temperature, we assumed instant inflaton fragmentation into radiation, the number of effective degrees of freedom is $g_* \approx 100$, and $T_\mathrm{CMB} \approx 2.7$ K is the present CMB temperature. First, in the $\tanh^2$ and similar models, the period scales as $T \sim \mth^{-1}\phi_0^{-1/2}$ (assuming $\phi_0 \gtrsim 10^{-4}$). Then, with $U_\text{preh}\approx U_0$ we obtain that the frequency $\fgw \approx T_\mathrm{CMB} \, \mth^{1/2} \approx 8\times 10^8$~Hz, which is out of the range of GW interferometers. Since this only depends on $\mth$, the CMB constraint on $A_s$ fixes $\fgw$. 

In contrast, the tachyonicity scale is not constrained in the delta model, and we can decrease $\fgw \sim T^{-1} \sim \Gamma_0$ by making $\Gamma_0$ small. Since we require $T\ll H^{-1}$, lowering $H$ (and $r$) will lower the bound on $\fgw$ and vice versa. In detail, we find the lower bound
\be
     \fgw \gg 5 \times 10^{-7} {\rm Hz} \times \left[ \frac{T_{\rm reh}}{\GeV}\right]\, .
\ee
For example, for $T_{\rm reh} = 1 \, \text{TeV}$, we find $\fgw \gg 0.5$ mHz, while at the current observational upper limit $T_{\rm reh} \approx 5 \times 10^{15} \GeV$, we find $\fgw \gg 2 \times 10^9$ Hz. 
Therefore, tachyonic preheating can generate GWs within the frequency range of LIGO-Virgo-Kagra \cite{LIGOScientific:2014pky, VIRGO:2014yos, Somiya:2011np} and future GW interferometers such as ET~\cite{Punturo:2010zz} and LISA~\cite{LISA:2017pwj} if the scale of inflation is sufficiently low.

\paragraph{Broad vs narrow resonances.}
Finally, let us compare our results to prior literature. It is worth noting that the band structure in tachyonic preheating is quite different from the more commonly studied parametric resonance. Early on \cite{Kofman:1994rk,Kofman:1997yn}, parametric resonance models were divided into two categories: those with narrow and broad resonance. In narrow resonance, the mode functions evolve almost adiabatically, and the resonance consists of multiple weak, narrow bands located roughly at integer values of $kT/\pi$ that do not extend to $k=0$. This resembles the behaviour of the orthogonal fields in our multi-field setup in section~\ref{sec:multi_field}. In the broad resonance regime, the resonance is dominated by one dominant band starting from $k=0$ and extending to high values of $kT$. Broad resonance is usually observed in multi-field setups for fields coupled to the inflaton. Our tachyonic resonance shares features from both: the dominant band always starts at $k=0$ and reaches up to $k = 2\pi/T$. The subsequent secondary bands appear periodically in $kT/\pi$, and the lowest ones, still partly tachyonic, are broad and strong and can significantly affect the preheating dynamics. Bands in the UV tail are weak and narrow.

In \cite{Soda:2017dsu,Kitajima:2018zco,Fukunaga:2019unq}, the authors discussed a `flapping resonance' similar to ours, where the inflaton repeatedly passes through tachyonic regions. Using an effective parameter $\tq$, they classified the resulting spectra to `broad' ($\tq \ll 1$), `intermediate' ($\tq \simeq 1$), and `narrow' ($\tq \ll 1$) \cite{Fukunaga:2019unq}, mimicking the classification discussed in the previous paragraph. In particular, they found that tachyonicity produces an intermediate value of $\tq$. However, the value of $\tq$ approaches zero for our $\tanh^2$ model in the limit of high $\phi_\amp$, for instance, $\tq = 0.034$ for $\phi_\amp = 7\phi_0$, even though these models do not have narrow resonances. At the same time, for the delta model, formally $\tq=\infty$, even though the resonance is not really of the broad form. This shows that the classification scheme of \cite{Fukunaga:2019unq} is not universal. The difference between our results arises from the length of the tachyonic period: in the models considered in \cite{Fukunaga:2019unq}, the tachyonic region was relatively short, whereas, in our models, the modes in the leading instability band are tachyonic for most of the time.

\section{Conclusions}
\label{sec:concl}

In this paper, we studied the linear regime of tachyonic preheating. The considered scenarios take place if the inflaton repeatedly returns to a tachyonic region of its potential, \eg, a plateau during post-inflationary oscillations. The presence of a tachyonic instability leads to a rapid fragmentation of the coherent background field within a fraction of an $e$-fold. Complementing our recent numerical study~\cite{Tomberg:2021bll}, we constructed simplified models by postulating the time-dependence of the effective mass of inflation perturbations. With these simplifications, the Floquet exponents determining the growth of each mode can be found analytically. As an applied example, we use these models to study tachyonic preheating for a physically well-motivated $\tanh^2$ potential.

As for general inflationary potentials leading to tachyonic preheating, the analytic models assume that the effective mass alternates periodically between tachyonic and non-tachyonic phases. We considered the following cases:
\begin{itemize}
    \item In the minimal simplified model---\emph{the delta model}---we neglected the duration of the non-tachyonic phase and described the effective mass squared by a delta function and a negative constant. Mode growth is characterized by two parameters: the period of oscillations and the constant negative mass squared. We showed that this ansatz is equivalent to assuming the idealized two-parameter potential $-m^2(|\phi| - \phi_0)^{2}/2$.
    
    \item As a generalization of the delta model, we considered \emph{the box model} which also accounts for the duration of the tachyonic period. In this set-up, mode growth is determined by four parameters: the constant effective masses and the durations of the tachyonic and non-tachyonic epochs. Motivated by numerical observations of the oscillating behaviour of realistic potentials, we further allow for a drop in the effective mass squared during transitions between the tachyonic and non-tachyonic phases, which we model by a delta function. As in the delta model, the postulated time dependence of the effective mass can be realized by an idealized piecewise quadratic continuous potential. The box model reduces to the delta model in the limit when the duration of the non-tachyonic period approaches zero.
\end{itemize}
In both models, the ans\"atze for the time-dependence of the effective mass squared contains an extra parameter related to the others by the requirement that the zero momentum mode must not grow.

Comparing the idealized analytic spectra of growth rates $\mu_k$ to numerical $\tanh^2$ results, we find that both models capture well the leading tachyonic peak in $\mu_k$. This justifies the usefulness of this simplification, as, in tachyonic preheating, the overall growth of the energy density, as well as the spectrum of the fragmented field component, is determined chiefly by the first instability band. The delta model, in particular, is useful for estimating the rate of tachyonic preheating since it provides an excellent analytic model of the first peak. The subleading peaks are better approximated by the box model. All in all, the simplified models provide a good analytical understanding of the band structure in tachyonic preheating. The most significant discrepancy with realistic models was found in the UV, where, for smooth potentials, the instability bands get exponentially narrower and weaker, while, for the simplified models, we observe only a power-law reduction with growing wavenumber. This discrepancy can be traced back to the discontinuities introduced by hand. This observation suggests that one can construct simplified models with improved UV behaviour by using ans\"atze with continuous higher time derivatives.

We also considered preheating in multi-field scenarios in the example of an $O(N)$ inflaton. In such scenarios, one direction of the field space corresponds to the inflaton degree of freedom while the rest behave as spectator fields. They are excited during preheating due to their coupling to the inflaton. We showed that only the inflaton is subject to a tachyonic instability as long as the potential is monotonous in $|\phi|$. The perturbations of the spectator components grow due to a less effective parametric resonance. Therefore, in the linear regime, the inflaton fragments into itself and the expected equipartitioning of the energy density among all components of the $O(N)$ multiplet completes during the subsequent non-linear evolution.

Based on our results, we make the following general observations on tachyonic preheating:
\begin{itemize}
    \item In typical tachyonic inflationary potentials, tachyonic preheating is effective if the field excursions 
    are sub-Planckian.
    
    \item Tachyonic preheating is expected to complete within less than an $e$-fold in all cases. Comparing the characteristic timescales $1/H = \sqrt{3/\rho}$, $1/\mu_\pk$ and $T$, we find that $1/H \gg 1/\mu_\pk > T$ in the tachyonic regime---damping inflaton oscillation amplitude is dominated by fragmentation/instability and the inflaton will oscillate at least a few times before fragmentation.

    \item Effective tachyonic preheating can be possible for any set of experimentally allowed inflationary parameters. In particular, it does not imply an extremely small tensor-to-scalar ratio $r$.
    
    \item The frequency of GWs produced during tachyonic preheating can range from nHz to GHz and may thus be accessible to future GW experiments.
\end{itemize}
These results can be used as guidelines for future model building with tachyonic preheating.

\acknowledgments

This work was supported by the Estonian Research Council grants PRG803, PRG1055, MOBTP135, MOBJD381 and MOBTT5 and by the EU through the European Regional Development Fund CoE program TK133 ``The Dark Side of the Universe."

\appendix
\section{Infrared stability of {\it O(N)} field perturbations}
\label{app:SON}

In this appendix, we will consider the infrared behaviour of the Floquet exponents in the scenario in which the inflaton belongs to a multiplet with an $O(N)$ global symmetry. At $k=0$, the equations for the inflaton mode~\eqref{eq:dphi_eom_flat} and the orthogonal modes~\eqref{eq:mode_perp} can be recast as $\delta \ddot \phi_{1,0} - (\dddot{\phib}_1/\dot{\phib}_1) \delta \ddot \phi_{1,0}=0$ and $\delta \ddot \phi_{\perp,0} - (\ddot{\phib}_1/\phib_1) \delta \ddot \phi_{\perp,0}=0$, respectively. Their two independent solutions read
\be\label{eq:u12,k=0}
    u_1 = 
\left\{\begin{array}{ll}
    \dot{\phib}_1   &  \qquad \mbox{(inflaton)} \\
    \phib_1         &  \qquad \mbox{(orthogonal)} 
\end{array}\right.,    
    \qquad
    u_2 = u_1 \int \frac{\td t}{u_1^2} \, .
\ee
These solutions have a unit Wronskian. 

In both cases, we can choose the point $t=0$ so that $\dot u_1(0) = 0$ without of loss of generality: for the inflaton modes, we choose $t=0$ when the inflaton crosses the minimum of the potential, so that $\dot u_1(0) = \ddot{\phib}_1(0) = 0$, and for the orthogonal modes, $t=0$ corresponds to a turning point since $\dot u_1(0) = \dot{\phib}_1(0) = 0$. In both cases, the $u_2$ integrand in \eqref{eq:u12,k=0} is singular at $T/2$, since $u_1(T/2) = 0$. For the inflaton modes, this corresponds to a turning point with $u_1(T/2) = \dot{\phib}_1(T/2) = 0$, while for orthogonal modes we have a zero-crossing, \ie, $u_1(T/2) = \phib_1(T/2) = 0$. Notice that here we use the assumption that the potential is symmetric---otherwise the field at the half period may not lie at the origin.

To address this singularity we follow the derivation in Ref.~\cite{Tomberg:2021bll} and construct the $u_2$ solution as~\cite{Tomberg:2021bll}
\be\label{eq:u2_piecewise}
    u_2(t) =
\left\{\begin{array}{lc}
    \tilde u_2(t)  &, \quad            0 \leq t \leq T/2            \\
    \tilde u_2(T-t) + \beta u_1(t) &, \quad   T/2 \leq t \leq T   
\end{array}\right. \, ,
\ee
with $\tilde u_2(t)$ defined by~\eqref{eq:u12,k=0} when $0 \leq t \leq T/2$. In this construction, $u_2$ is continuous while the continuity of $\dot u_2$ gives
\bea
    \beta 
&   = \frac{2\dot{\tilde u}_2(T/2)}{\dot u_1(t)}
    = \lim_{t\to T/2} \frac{2}{\dot{u}_1(T/2)} \left[ \dot{u}_1(t) \int^t_{0} \frac{\td t'}{u_1(t')^2} + \frac{1}{u_1(t)} \right]
\eea
so that the monodromy matrix around $k=0$ reads
\bea
    \frac{1}{2}\Tr G 
    &= - 1 + \frac{k^2}{2} \beta \int^T_0 \td t'\, u_1(t')^2 + \mathcal{O}(k^4).
\eea
The derivation is similar for asymmetric potentials, although slightly more tedious. In the latter case, one also has $\Tr G/2 = 1$ instead of $\Tr G/2 = -1$ at $k=0$.

Plugging in $u_1$ from Eq.~\eqref{eq:u12,k=0} and eliminating the derivatives using the equations of motion and energy conservation for the background field, we find that
\be
    \frac{1}{2}\Tr G
    = - 1 - \frac{k^2}{2} W \partial_{\rhob} T  + \mathcal{O}(k^4)
\ee
for the inflaton modes. The abbreviated action $W$ was defined in Eq.~\eqref{eq:W}. For the orthogonal modes
\bea
    \frac{1}{2}\Tr G_{\perp} 
    &= - 1 + k^2 \frac{\beta_{\perp}}{\sqrt{2\rhob}} \int^{\phi_{\rm amp}}_0 \frac{\td \phi\, \phi^2}{\sqrt{1 - V(\phi)/\rhob }} + \mathcal{O}(k^4)\, ,
\eea
where $\rhob = V(\phi_{\rm amp})$ and $\beta_\perp$ does not have a short analytical expression. However, since $\beta_\perp \geq 0$, we have that $|\Tr G_{\perp}| \leq 1$ around $k=0$ and thus there must exist a $k_1$ such that $k \in [0,k_1]$ is stable. In other words, the $\mu_k$ spectrum for orthogonal perturbations begins with a stability band. The instability bands of orthogonal modes thus resemble a narrow parametric resonance.

\bibliography{refs_preheating.bib}
\end{document}